\documentclass[11pt]{book}
\usepackage{Wiley-AuthoringTemplate}
\usepackage[numbers]{natbib}

\usepackage{multirow}

\usepackage{glossaries}
\usepackage{subfigure}

\DeclareMathOperator*{\maximize}{maximize} 
\DeclareMathOperator*{\minimize}{minimize} 
\DeclareMathOperator*{\argmin}{argmin} 
\DeclareMathOperator*{\subjectto}{subject \hspace{3pt} to:\hspace{3pt}} 
\makeindex

\begin{document}
\setcounter{chapter}{13}
\chapter[Machine Learning for Metasurfaces]{Machine Learning for Metasurfaces Design and Their Applications\protect\footnote{K. V. M. acknowledges support from the National Academies of Sciences, Engineering, and Medicine via Army Research Laboratory Harry Diamond Distinguished Postdoctoral Fellowship.}}

\author*[1]{Kumar Vijay Mishra}
\author[2]{Ahmet M. Elbir}
\author[1,3]{Amir I. Zaghloul}

\address[1]{\orgdiv{Computational and Information Sciences Directorate (CISD)}, 
\orgname{United States DEVCOM Army Research Laboratory}, 
\postcode{20783}, \countrypart{Adelphi}, 
     \city{Maryland}, \country{USA}}%

\address[2]{\orgdiv{Interdisciplinary Centre for Security, Reliability and Trust (SnT)}, 
\orgname{University of Luxembourg}, 
\postcode{1855}, \street{Av. John F. Kennedy}, \country{Luxembourg}}%

\address[3]{\orgdiv{Bradley Department of Electrical and Computer Engineering}, 
\orgname{Virginia Tech}, 
\postcode{24061}, \countrypart{Virginia}, 
     \city{Blacksburg}, \country{USA}}%
     
\address*{Corresponding Author: Kumar Vijay Mishra; \email{kvm@ieee.org}}

\maketitle

\begin{abstract}{Abstract}
Metasurfaces (MTSs) are increasingly emerging as enabling technologies to meet the demands for multi-functional, small form-factor, efficient, reconfigurable, tunable, and low-cost radio-frequency (RF) components because of their ability to manipulate waves in a sub-wavelength thickness through modified boundary conditions. They enable the design of reconfigurable intelligent surfaces (RISs) for adaptable wireless channels and smart radio environments, wherein the inherently stochastic nature of the wireless environment is transformed into a programmable propagation channel. In particular, space-limited RF applications, such as communications and radar, that have strict radiation requirements are currently being investigated for potential RIS deployment. The RIS comprises sub-wavelength units or meta-atoms, which are independently controlled and whose geometry and material determine the spectral response of the RIS. Conventionally, designing RIS to yield the desired EM response requires trial and error by iteratively investigating a large possibility of various geometries and materials through thousands of full-wave EM simulations. In this context, machine/deep learning (ML/DL) techniques are proving critical in reducing the computational cost and time of RIS inverse design. Instead of explicitly solving Maxwell's equations, DL models learn physics-based relationships through supervised training data. The ML/DL techniques also aid in RIS deployment for numerous wireless applications, which requires dealing with multiple channel links between the base station (BS) and the users. As a result, the BS and RIS beamformers require a joint design, wherein the RIS elements must be rapidly reconfigured. The lower computation time and model-free nature of DL make it robust against data imperfections and environmental changes in RIS-aided communications. At the physical layer, DL has been shown to be effective for RIS signal detection, channel estimation, and active/passive beamforming using architectures such as supervised, unsupervised, and reinforcement learning. This chapter provides a synopsis of DL techniques for both inverse RIS design and RIS-assisted wireless systems. 
\end{abstract}

\keywords{Beamforming, deep learning, inverse design, metasurfaces, smart radio environment.}

	\section{Introduction}
	\label{sec:Introduciton}
	The emerging industrial use-cases of sixth-generation (6G) and beyond wireless networks are envisaged to include industrial automation, autonomous vehicles, and smart infrastructure. These applications require significant improvements in data capacity, system latency, and quality-of-service reliability over the current 5G networks. In this context, \textit{reconfigurable intelligent surface} (RIS) has been identified as a key enabling technology to program the \textit{smart radio environment} (SRE), increase link quality, and reduce the hardware complexity \citep{renzo2019smart,hodge2020intelligent}. The RIS is made up of a \textit{metasurface} (MTS) - a two-dimensional (2-D) reconfigurable electromagnetic (EM) layer composed of a large periodic array of subwavelength scattering elements (meta-atoms) with specially designed spatial features \citep{holloway2012overview,nguyen2019retrieval}. Compared to electrically large arrays, the nearly passive meta-atoms offer lower cost and power consumption. The radio-frequency (RF) MTS performs customized transformations, such as beamforming, on a reflected incident wave through modified surface boundary conditions using Huygens’ principle. For example, the MTS shifts the reflected phase of incident signal by creating a field discontinuity at the boundary of the surface. The arrangement and subwavelength structure of each meta-atom and, in turn, the array of space- and time-varying meta-atoms determine MTS aperture field distribution and control the direction and strength of reflected signal \citep{hodge2019joint}. 
	
	In a conventional wireless communication systems, the network optimization has been limited to control at the transmitter and receiver. This paradigm assumes that the wireless fading channel is uncontrollable and is a significant factor limiting the performance because of random signal reflections, diffraction, and scattering in the wireless environment. The RIS overcomes many of the aforementioned fading channel limitations through the ability of MTS to manipulate waves, achieve arbitrary aperture beamforming, and perform real-time analog spatial signal processing. This has spawned novel MTS-based RF applications such as intelligent beamforming \citep{wu2019towards}, anomalous refraction and reflection \citep{chen2018huygens}, frequency selective and high-impedance surfaces \citep{sievenpiper1999high}, scattering reduction \citep{su2017ultra}, polarization conversion \citep{zhu2013linear}, leaky-wave antenna \citep{minatti2015modulated}, 
surface wave control \citep{maci2011metasurfing}, beam focusing \citep{mishra2018reconfigurable}, transmit-array antennas \citep{li2015x}, reflect-array antennas \citep{chen2016review}, and holographic imaging \citep{glybovski2016metasurfaces}. Initial applications of RIS were limited to wireless communications for interference suppression \citep{wu2019towards}, joint wireless information and power transmission \citep{kumar2022feasibility}, physical layer security \citep{mishra2022optm3sec}, and multi-beam design \citep{torkzaban2021shaping}. However, more recent works have introduced IRS to radar remote sensing \citep{esmaeilbeig2022irs,wang2022stars,esmaeilbeig2022cramer} and joint radar-communications systems~\citep{wei2022irs,elbir2022rise}.

\begin{figure}
		\centering
		{\includegraphics[width=1.0\textwidth]{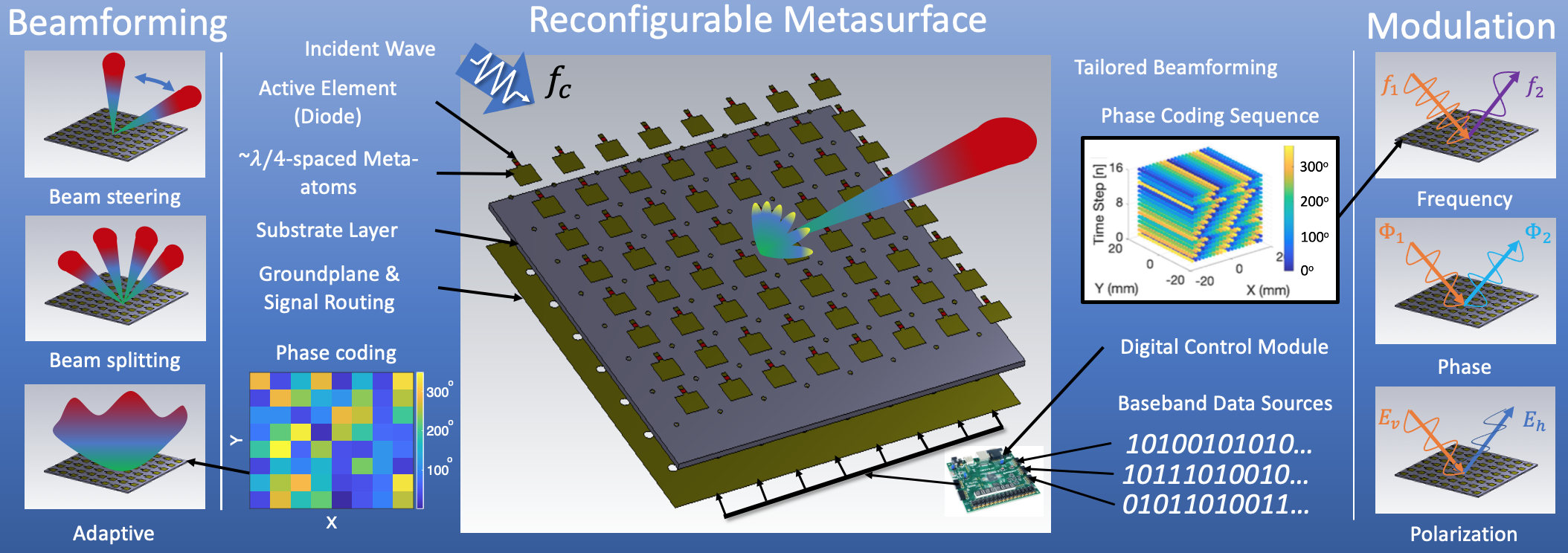}}
		\caption{The RIS architecture (center) operating at carrier frequency $f_c$ for wireless communication networks comprises meta-atoms located at below carrier wavelength ($\lambda$) spacing. It acts as both an endpoint transceiver and a relay. The RIS enables various beamforming functions (left column) including beam steering, splitting, and adaptive beamforming for customized radiation patterns by manipulating the phase coding of constituent meta-atoms. It is also capable of directly modulating (right column) the surface in frequency, phase, and polarization \citep{hodge2021deep}.
		}	
		\label{fig_overview}
	\end{figure}

    In a wireless link, the RIS functions as either an electrically large antenna array at the endpoints or as an amplify-and-forward relay (Fig. \ref{fig_overview}). By actively controlling and optimizing the amplitude/phase of each meta-atom across the aperture, the RIS maximizes the receive signal-to-noise ratio and provides adaptive beamforming to coherently focus the reflected signal on the receiver. Through joint optimization of the wireless channel and endpoints, RIS-assisted links are able to realize SRE. 
    Each scattering element typically includes an active tuning element, such as a varactor or PIN diode, whose bias voltage is software-controlled to change the EM response of the surface. The bias voltage for each meta-atom is pre-computed and modulated by a digital control module employing a field programmable gate array (FPGA) \citep{hodge2019reconfigurable}. Each meta-atom is controlled by tuning its EM properties (susceptibility or impedance) which affects the spectral response of the reflected signal. This aids in producing tailored radiation patterns for diverse functions, such as beam steering, anomalous reflection, focusing, beam splitting, absorption, and direct modulation of the reflected signal.  

There are several challenges in the design, fabrication, deployment, and processing of RIS. In applications such as radar and communications that have precise radiation pattern constraints, the RIS design often involves optimization of several complicated and irregular geometry parameters to meet the required resonant frequency, gain, polarization, bandwidth, and size constraints. The conventional design process could be very tedious. Further, post-deployment, the processing of RIS signals and optimized beamforming is also challenging because of high-dimensional nature arising from the use of several antennas. In this context, machine learning (ML) techniques have recently shown unprecedented performance in problems where it is challenging to develop an accurate mathematical model for feature representation. These methods are now also transforming the above-mentioned tedious approaches to design RIS and process its signals. In particular, as a class of machine learning techniques, deep leaning (DL) methods have gained much interest recently for solving many challenging problems such as speech recognition, visual object recognition, rainfall estimation, and language processing \cite{deepLearningScience, deppLearningRepresetation,deepLearning4SignalProcessing}. These techniques offer advantages such as low computational complexity while solving optimization-based or combinatorial search problems as well as the ability to extrapolate new features from a limited set of features contained in a training set \cite{deepLearningScience}. Recently, DL for MTS inverse design, wherein a meta-atom design if synthesized from a specific response, has become very popular. This has been applied for semi-automated inverse design of metamaterials \cite{ma2018deep}, MTS \cite{zhang2018machine,qiu2019deep}, and nanophotonic structures \cite{peurifoy2018nanophotonic}. Note that the above ML/DL application to MTS/RIS design is different from using DL to perform signal processing functions in RIS-aided communications (see, e.g., \citep{elbir2020survey} for a survey). In the following, we describe these aspects in detail.

\subsection{ML/DL for RIS Design}
The design and optimization of RIS hardware at the physical layer remains a formidable challenge. To date, RIS/MTS implementations remain quite limited. To realize the promise of RIS-assisted networks and SRE, more robust and automated MTS design techniques are required. Without capable RIS hardware, the benefit of RIS-assisted networks will be significantly reduced due to EM limitations. In general, canonical structures such as v-antennas, loaded-dipoles, split-ring resonators, are used to fabricate RIS. However, meta-atoms based on these geometries usually fall short of desired performance, particularly when anisotropic, broadband, and/or wide-angle responses are required. As a result, traditional MTS design approaches exhibit performance limitations, especially given the complexity of MTS hardware requirements and increasing functionality required for wireless nodes in next generation networks. 
 
 Designing a user-defined, arbitrary wave-front RIS or \textit{metagrating}  \cite{liu2018generative,jiang2018data,hodge2019rf} is a challenging, labor-intensive, and long process. In general, a new MTS design entails numerous rounds of manual tuning and full-wave simulations that iteratively solve Maxwell’s equations until a locally optimized design is achieved \citep{liu2018generative}. Initial designs are typically based on physical instincts and intuitive arguments. However, the final geometric structure and material characteristics are attained through iterative analyses.
 
The ML/DL approaches expend computational time and resources upfront as a fixed-cost to generate training data sets of device geometries and their associated spectral responses but are useful during the predication stage \cite{campbell2019review}. Deep neural networks are trained to map the nonlinear relationships between meta-atom geometry and spectral response. The power of deep neural networks comes from their multi-layered composition which allows them to learn the relationships between data with multiple levels of abstraction \cite{lecun2015deep}. Once trained, a deep neural network efficiently produces the geometry of a meta-atom given a desired spectral response. The application of deep learning to the inverse design of MTS and nanophotonic structures is still in its early stages and much more work required to realize more generalized complex designs, reduce the amount of required training data, and result in increased efficacy. Nearly all of these works rely on supervised learning techniques for metamaterial performance predictions, which map known input-output pairs based on large training examples. In MTS design, applying such techniques does not result in new shapes different than the ones used in training. This severely limits the ability to generate customized MTS patterns. In \citep{hodge2019rf, hodge2019joint}, we introduced the use of generative adversarial networks (GANs) to microwave MTS design that aids in discovering new shapes of meta-atoms. 

\subsection{ML/DL for RIS Applications}
\begin{figure}
		\centering
		{\includegraphics[width=0.8\textwidth]{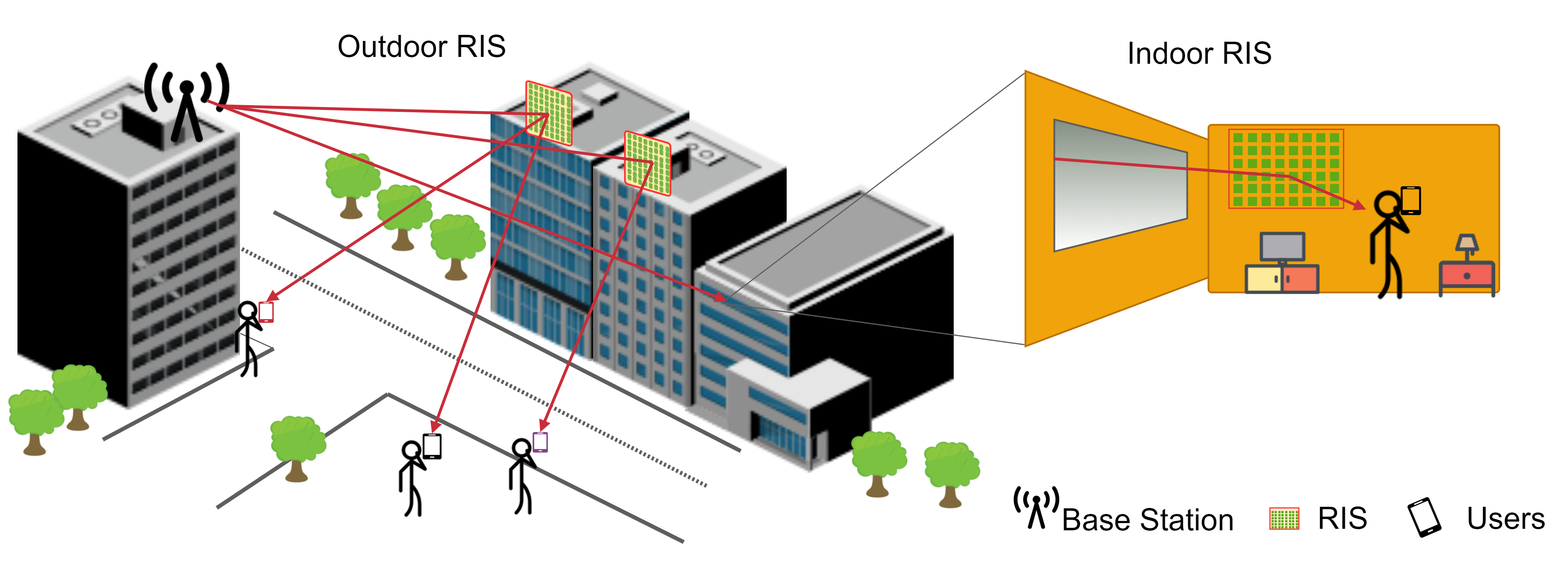} }
		\caption{RIS-assisted wireless communications for outdoor and indoor deployments. A BS on top of the infrastructure (left) communicates with the users on ground through an intermediate RIS mounted on other buildings (center). The BS also serves users (right) inside the apartment building through an RIS placed on the wall of the room \citep{elbir2020survey}. 
		}
		\label{fig_IRS}
	\end{figure}

The next-generation millimeter wave (mm-Wave) massive multiple-input multiple-output (MIMO) systems require large antenna arrays with a dedicated radio-frequency (RF) chain for each antenna. This results in expensive and large system architectures which consume high power and processing resources. To reduce the number of RF chains while also maintaining sufficient beamforming gains, hybrid analog and digital beamforming architectures were introduced. However, the resulting cost and energy overheads using these systems remain a concern. 
Recently, RISs have emerged as a feasible solution~\citep{irs_survey1} to implement low cost and light-weight alternative to large arrays complexity in both outdoor and indoor applications, usually with separate operating frequencies or spectral bands. (Fig.~\ref{fig_IRS}).
	
The RISs reflect the incoming signal by introducing a pre-determined phase shift. This phase shift is controlled via external signals by the base station (BS) through a backhaul control link. As a result, the incoming signal from the BS can be manipulated in real-time, thereby, reflecting the received signal toward the users. Hence, the usage of RIS enhances the signal energy received by distant users and expands the coverage of the BS. It is, therefore, required to jointly design the beamformer parameters both at the RIS and BS. This achieves desired channel conditions, wherein the BS conveys the information to multiple users through the RIS~\citep{irs_RL_BF_IRSonly}.  Different from amplify-and-forward (AF) relay systems, an  RIS can have both active and passive components, which can provide a flexible configuration, thus, it has less active transmit modules or totally reflects the received signal as a passive surface. Thus, the RIS is much more energy- and  spectrum-efficient~\citep{irs_Indoor_locationToBF}.

	The accuracy of beamformer design strongly relies on the knowledge of the channel information. In fact, the RIS-assisted systems include multiple communications links, i.e., a direct channel from BS to users and a cascaded channel from BS to users through RIS. This makes the RIS scenario even more challenging than the conventional massive MIMO systems. Furthermore, the wireless channel is dynamic and uncertain because of changing RIS configurations. Consequently, there exists an inherit uncertainty stemming from the RIS configuration and the channel dynamics. These characteristics of RIS make the system design very challenging~\citep{irs_RL_energyEfficient_,irs_RL_BF_IRSonly}.
	
	To address the aforementioned uncertainties and non-linearities imposed by channel equalization, hardware impairments, and sub-optimality of high-dimensional problems, model-free techniques have become common in wireless communications~\citep{dl_WCM}. In this context, DL is particularly powerful in extracting the features from the raw data and providing a ``meaning'' to the input by constructing a model-free data mapping with huge number of learnable parameters. Furthermore, DL is helpful when modeling the channel characteristics thanks to its data-driven structure. 
		A learning model constructs a non-linear mapping between the raw input data and the desired output to approximate a problem from a model-free perspective \citep{dl_WCM}. Thus, its prediction performance is robust against the corruptions/imperfections in the wireless channel data. 
	DL learns the feature patterns, which are easily updated for the new data and adapted to environmental changes. In the long run, this results in in lower computational complexity than a model-based optimization~\citep{irs_RL_BF_IRSonly}. 
		DL-based solutions have significantly reduced run-times because of parallel processing capabilities. On the other hand, it is not straightforward to achieve parallel implementations of conventional optimization and signal processing algorithms \citep{elbir2020_FL_CE}. 
	The aforementioned advantages have led to DL superseding the optimization-based techniques in the RIS system design for physical layer of the wireless communications~\citep{dl_WCM}. 

\subsection{Organization}
This chapter provides an overview of recent developments in using ML/DL for designing, deploying and processing the physical layer of RIS. The rest of the chapter is organized as follows. In the next section, we discuss various ML techniques for inverse RIS design. Then, we introduce various techniques DL for RIS design in \ref{sec:dl_inverse} and provide a few case studies in  Section~\ref{sec:case_studies}. Then, we focus on DL-aided RIS applications for wireless systems in Section~\ref{sec:app}. including signal detection and channel estimation. For a more widely used application of RIS beamforming, we discuss various DL frameworks in Section~\ref{sec:beamforming}. We also discuss current challenges in using ML/DL for RIS systems and highlight related future research directions in Section~\ref{sec:pathForward}. We conclude in Section~\ref{sec:summ}.

\section{Inverse RIS Design}
\label{sec:ris_design}
\begin{figure}
    \centering
    \includegraphics[width=1.0\textwidth]{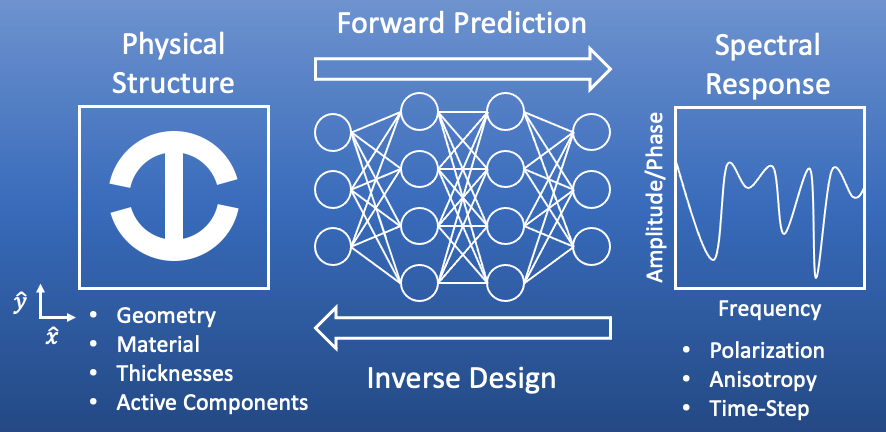}
    \caption{In inverse RIS design, ML algorithms learn and generalize complex EM relationships between the physical RIS structure (left column) and spectral response (right column) through training data \citep{hodge2021deep}. 
    }
    \label{fig:MLexample}
    \end{figure}

	Communications-based analysis of RIS without physics-based EM-compliant models is a major limitation of current research. 
	Until recently, prior works did not consider such realistic RIS implementations. As the parameter spaces of meta-atom geometry and constituent materials has grown, the conventional approaches to achieve the targeted EM response have become more tedious. In this context, learning models have demonstrated the ability to implicitly learn Maxwell's equations from training data within a constrained design space. The ML techniques have witnessed increased use in research to create surrogate models for MTS performance prediction, inverse design, and optimization. For an inverse MTS design problem, the input is an arbitrary design spectrum and the network finds or synthesizes a geometry to closely approximate the desired spectral response (Fig.~\ref{fig:MLexample}). 
	
	
	Major benefits of DL-based RIS design for wireless communications include:
	\begin{itemize}
    \item \textit{EM-based surrogate models}: DL constructs a nonlinear mapping between the raw input data (meta-atom design) and the desired output to approximate the MTS response.
    \item \textit{Inverse design}: Deep generative models are utilized to learn geometric features from training data and generate new meta-atom designs to achieve the spectral response.
    \item \textit{Diverse EM surface representations}: DL-based MTS design admits flexible design representation. The input could be either vectors of discrete parameters describing the geometry, material, frequency, and angular design parameters or pixelated images to represent the geometry or phases of the meta-atom design. Whereas a fully-connected neural network is well-suited to process the simple designs specified by the former representation, a convolutional networks handle images appropriately to yield more complex MTS geometries.
    \end{itemize}

\begin{table}
\centering
\caption{State-of-the-art on MTS inverse design \citep{hodge2021deep}
\label{tab:my_table}}
{\tiny
\begin{tabular}{|| p{0.13\linewidth} || p{0.12\linewidth} | p{0.08\linewidth} | p{0.15\linewidth} | p{0.31\linewidth} | p{0.23\linewidth} ||}
\hline
  \textbf{\scriptsize{Algorithm}} &  
  \textbf{\scriptsize{Frequency}} &
  \textbf{\scriptsize{MTS layers}} &
  \textbf{\scriptsize{Data}} &
  \textbf{\scriptsize{Key features}} &
  \textbf{\scriptsize{Drawbacks}} 
  \\ 
  \hline
\multicolumn{6}{|c|}{\textbf{Evolutionary optimization techniques}} \\ \hline 
GA \citep{campbell2019review}        &  $15$-$45$ GHz & $1$ & Parameter Vector & Pixelized meta-atoms with discrete input design space when a contiguous structure is not required & Optimization from scratch for each design; output structures may be too complex to fabricate \\ \hline
PSO \citep{campbell2019review}       & $9.5$-$12$ GHz & $1$  & Binary Matrix (2-D) & Swarm-based GO technique for pixelized meta-atom design; outperforms GA for various EM designs & Optimization from scratch for each design with parameter tuning \\ \hline
ACO \citep{campbell2019review}       & $1$-$4.5$ GHz & $3$ & Binary Matrix (3-D) & MTS, including 3-D structures and wire grid arrays, with discrete design space and a contiguous structure & Optimization from scratch for each design; output structures may be too complex to fabricate \\ \hline
\multicolumn{6}{|c|}{\textbf{Learning methods}} \\ \hline 
ANN \citep{peurifoy2018nanophotonic} & $375$-$749$ THz & $1$ & Parameter Vector & Performance prediction, inverse design, and optimization of nanophotonic particles & Limited design variables; applicable to only spherical dielectric nanoparticles \\ \hline
ANN \citep{inampudi2018neural}       & $193$ THz       & $1$ & Parameter Vector & Performance prediction and inverse design of metagratings & Limited set of parametric inputs; significant training overload  \\ \hline
DNN \citep{ma2018deep}               & $30$-$80$ THz   & $2$ &  Parameter Vector & Inverse design of chiral and multi-layer MTS & Design-specific architecture; 
limited design space \\ \hline
CNN \citep{zhang2019machine}         & $10$ GHz        & $1$ & Binary Matrix (2-D) & Anisotropic digital coding MTS; PSO for beamforming & Significant training overload  \\ \hline
CNN \citep{shan2020coding}           & $9.37$ GHz      & $1$ & Binary Matrix (2-D) & Hybrid CNN-GA for space-time modulation of programmable MTS; multi-beam steering & Binary phase coding limits beamforming performance; limited tunability \\ \hline
cDC-GAN \citep{liu2018generative}    & $170$-$600$ THz & $1$ & Image Matrix (2-D) & Generative inverse design of transmission MTS & Significant training overload; limited to single layer designs and passive structures \\ \hline
cDC-GAN \citep{hodge2019rf}          & $1$-$30$ GHz    & $1$ & Image (2-D) & Reflective RF MTS; training set with published meta-atom structures to improve learning  & Limited to single layer; post-processing required \\ \hline
cDC-GAN \citep{hodge2019joint}       & $5$-$25$ GHz    & $2$ & Image (3-D) & Multi-layer MTS; RGB-style matrix to represent multiple layers & No active elements; additional validation required \\ \hline
cDC-GAN \citep{hodge2019multi}       & $5$-$25$ GHz    & $3$ & Image (3-D) & Federated learning for multi-layer design & Significant training overload \\ \hline
cDC-VAE \citep{ma2019probabilistic}      & $40$-$100$ THz  & 1 & Image (2-D) & Anisotropic MTS; encodes input into low-dimensional latent space & 
Significant training overload; post-processing required
\\ \hline
TO-GAN \citep{jiang2019free}             & $231$-$600$ THz & 1 & Image (2-D) & Free-form diffractive metagrating design for select wavelength-deflection angle pairs with topology refinement & Additional optimization required 
\\ \hline
GLOnet \citep{jiang2019global}       & $231$-$500$ THz & 1 & Image (2-D) & 
Dielectric MTS design without training sets & Limited to single objective optimization; requires solving Maxwell's equations inside training loop \\ \hline
\end{tabular}
 }{}
\end{table}\normalsize


Table~\ref{tab:my_table} summarizes prior works on various techniques for RIS inverse design. The non-DL methods typically comprise of several evolutionary optimization algorithms as listed below. The drawback of traditional optimization techniques is that they start from scratch with each new design. This often requires hundreds of additional full-wave simulations per design.
	\subsubsection{Genetic algorithm (GA)}	
	This is an iterative global optimization (GO) algorithm that has been used extensively in the design of pixelated coded MTS designs. GA is a nature-inspired algorithm that uses binary strings (chromosomes) to represent candidate designs \citep{campbell2019review}. During the optimization, the GA selects the best subset of design candidates from the previous generation to serve as starting points for mutation and crossover in the next design iteration. Recent GA applications include coding MTS \citep{campbell2019review} which demonstrates channel response modification, efficient polarization conversion, and phase-graded beam steering.
	
	\subsubsection{Particle swarm optimization (PSO)}
	A popular stochastic evolutionary computation technique, PSO is inspired by the movement and intelligence of swarms. Recently, it has been employed for shaping EM waves using pixelized coded metasurfaces \citep{campbell2019review}. The design procedure using PSO is tied to a full-wave EM solver and completely automatic. The software yields both microscopic meta-atom designs and the macroscopic aperture coding matrix. By changing the reflection phase difference between cells, this approach has produced designs of functional metasurfaces with circularly- and elliptically-shaped radiation beams and multi-beam patterns. This is useful for achieving customized radiation patterns to enhance link performance in the wireless communication channel. 
	Similar efforts have used a simulated annealing algorithm for the design and optimization of a broadband diffusion MTS using anisotropic elements for scattering reduction. In \citep{zhang2019machine}, binary PSO (BPSO) was used to automate the macroscropic layout of both passive and active aperture to realize user-defined dual-beam scattering radiation patterns. For example, this study used BPSO to realize a reflecting MTS with a left-handed circular polarization (LHCP) beam and a right-handed circular polarization (RHCP) beam. Results of this study were experimentally verified. This digital coding approach has been applied to both passive and active R-MTS.

	
	\subsubsection{Ant colony optimization (ACO)} This is another swarm-based algorithm inspired by \textit{stigmergy} in ant colonies in order to search for optimal solutions to graph-based problems \citep{campbell2019review}. Here, a number of \textit{artificial ants} build solutions to an optimization problem and exchange information on their quality using a cooperation scheme similar to that utilized by real ants. In \citep{campbell2019review}, inverse MTS design is performed based on multi-objective lazy ACO (MOLACO) to synthesize 3-D nano-antenna geometries with low-loss transmission performance and broad phase tunability. The ACO is generally most useful for a discrete input design space and when a contiguous structure is required.
	
	\section{DL-Based Inverse Design and Optimization}
	\label{sec:dl_inverse}
	The computational power and time required for evolutionary optimization algorithms grow exponentially with the number of design parameters. This is mitigated by DL-based inverse design for RIS. Prior works have employed a variety of network structures and algorithms based on the availability of data, RIS topology, and desired EM spectral response.
	\subsection{Artificial Neural Network (ANN)}
	\label{sec:ANNs}
    The ANNs were first used to approximate light scattering by multi-layer nanoparticles (meta-atoms) \citep{peurifoy2018nanophotonic}. Similar to MTS, nanophotonic particles derive their frequency response from physical structure and the size constituent scatterers. Then, \citep{inampudi2018neural} used a similar technique for metagratings. Typical inverse design problems require optimization in high-dimensional space, which involves lengthy calculations and are typically solved using genetic algorithm or adjoint methods. However, the computational power and time required for GA optimization grows exponentially with the number of design parameters. 
    
	The primary application of ANNs in MTS design is performance approximation. The feedforward ANN is trained to be a high-fidelity surrogate model for performance prediction. Using training data consisting of meta-atom physical design parameters as inputs and frequency response as labels, the ANN is trained to approximate a complex physics simulation (such as finite-element method (FEM), method of moment (MoM), or finite-difference time-domain (FDTD) simulation). Through the training data, the 
	ANN learns to map the scattering function of the meta-atom into a continuous, higher-order space where the derivative is found analytically through propagation. In \citep{peurifoy2018nanophotonic}, a trained ANN simulated spectral responses orders of magnitude faster than conventional full-wave simulations. This study used a fully-connected ANN consisting of four layers with $250$ neurons per layer resulting in 239,500 parameters. The inputs were the thickness of each meta-atom layer (the materials were fixed) and the outputs were the spectrum sampled at points between $400$ and $800$ nm. 
	The results suggest that the ANN was not simply fitting the data, but rather discovered the underlying structure of input-to-output mapping to generalize the physics of the systems with the training set and solve problems not yet encountered. 
	
	A significant drawback of this approach is that the inputs are limited to the thicknesses of the meta-atom layers with fixed materials. This results in a lack of generalizability for the ANN that vastly limits the possible meta-atom design structures. While fixing the input parameters reduces the complexity of the ANN architecture, it limits the design space and optimal designs. Another drawback of this approach is that \citep{peurifoy2018nanophotonic} required $50,000$ examples using conventional simulation methods to generate training data. However, unlike evolutionary optimization methods such as GA or PSO, simulation of the training dataset is an upfront fixed cost because it only needs to be simulated once and is then leveraged for other designs. Additionally, the simulations for training data generation are highly parallelized unlike serial optimization techniques.

	Once trained, \citep{peurifoy2018nanophotonic} shows that the ANN solves inverse design problems more quickly than than its numerical counterparts because the gradient is found analytically, through back propagation, rather than numerically. Similar to inverse design, the ANN also optimizes for a desired property by altering the cost function used for the design without training the ANN. Their results that the ANN performs inverse design and optimization more accurately than traditional numerical nonlinear optimization techniques.
	
	\begin{figure}
		\centering
		\includegraphics[width=1.0\textwidth]{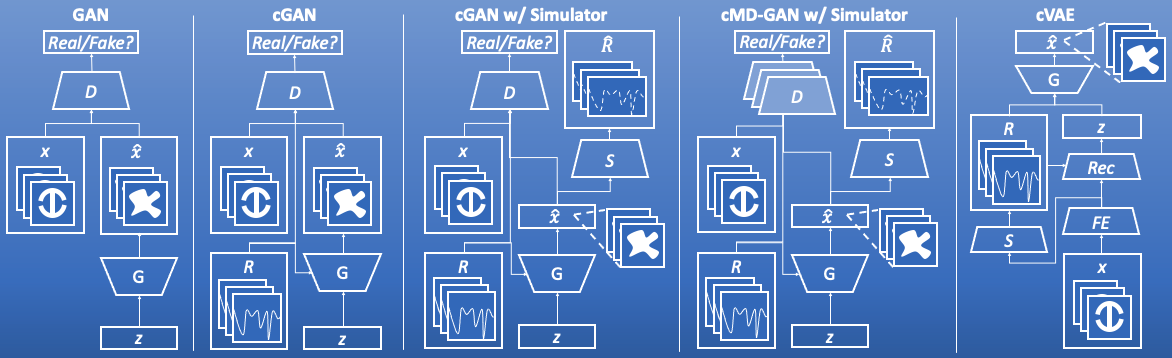}
		\caption{DGM architectures for RIS inverse design. The conventional GAN (left) lacks spectral information of the RIS structure $x$. The latent variable $z$ is fed to the generator $G$ to yield an estimated meta-atom structure $\hat{x}$. The discriminator $D$ then makes a decision if $\hat{x}$ is a valid design. The cGAN (second from left), conditioned by the reflection spectra $R$, shows improved performance. A simulator neural network may also be added to cGAN (center) to accelerate training and also predict the performance $\hat{R}$ of generated meta-atoms. The cMD-GAN (second from right) comprises of multiple discriminators, one for each layer. The cVAE (right) consists of an encoder-decoder network structure, where a feature extractor (FE) coupled with the recognition (Rec) network serves as the encoder to map the meta-atom structure to a lower-dimensional latent variable space. 
		The generation model is a reconstruction (decoder) network \citep{hodge2021deep}. 
		}
		\label{fig:netStructures}
	\end{figure}
	
	\subsubsection{Deep Neural Networks (DNN)}
	To model more complex meta-atom structures and increase performance prediction accuracy, DL has been applied to the on-demand design of chiral (a form of anisotropy) MTS \citep{ma2018deep}. Here, deep neural networks (DNN) - an ANN comprised of many hidden layers to significantly expand learning and generalization ability - was employed to automatically design and optimize 3-D chiral MTS with strong anisotropic spectra at predetermined wavelengths. The network comprised two bidirectional networks that were constructed using partial stacking technique. This study limited the input design space (and hence the structures obtained) and predicted the reflection spectral response at $201$ discrete frequency points for two orthogonal polarization and the cross-polarization coupling term resulting in a $3$-by-$201$ spectral output vector. By fixing the inputs to be five specific design parameters, this DNN design approach is also limited in its generaliziblity to other physical structures in the design space. 
	Full-wave simulation was used to generate the training data set for $30,000$ example meta-atoms. The DNN achieved high efficiency and high-accuracy for performance prediction and inverse design for anisotropic MTS, where the meta-atom design space is limited.
	
	\subsection{Convolutional Neural Networks (CNN)}
	\label{sec:CNNs}
	To improve on the lack of generalization and increase performance prediction accuracy, convolutional neural networks (CNN) are used to design anisotropic digital coding metasurfaces. CNNs are a class of ANNs that use convolution functions to learn hierarchical patterns within data. These models learn generalized patterns across many spatial scales from their input data and are widely used on image data. In \citep{zhang2019machine}, a CNN predicted the reflection phase response of binary coded meta-atoms where each meta-atom contains $16$-by-$16$ square sub-pixels and is mirrored with two-fold symmetry. The CNN used in this study is a $101$-layer deep residual network, known as Resnet-101. The authors found that other networks with fewer layers resulted in less precise and robust performance predictions. 
	
	The results show an accuracy of $90.05\%$ of phase responses with $2^{\circ}$ error in the $360^{\circ}$ phase. A drawback of this binary coding approach is that a $16$-by-$16$ pixel meta-atom has $2^{16}$ potential design combinations. This study generated training data by simulating randomized pixel matrices. However, it was fundamentally inefficient in an analogous manner to GA because the training data is essentially random and does not contain the knowledge of canonical structures in the training data set. This likely results in significantly more required training data and greater network complexity. Another drawback of this study is that it required full-wave simulation of 70,000 training examples 10,000 test examples to generate the training dataset.
	
	A significant CNN advantage is that the meta-atom shape is directly input into the network rather than shape-specific design parameters. The convolutional filters allow the CNN to learn the physical structure that leads to given EM response, leading to a broader applicability of the model.
	
	In \citep{shan2020coding}, the element phases of a reconfigurable MTS were computed by a 11-layer CNN for multiple beam steering applications. The input was the parameter vector representing the target beam pattern and the output was a matrix that carried the 1-bit codes for a programmable $2304$-element MTS. This technique to obtain the phase matrices reduced the time for producing almost similar beam patterns using conventional methods to a few milliseconds.

	\subsection{Deep Generative Models (DGMs)}
	Generative models are unsupervised or semi-supervised learning models that infer a function to describe hidden structure from unlabeled data. Their functions include clustering, density estimation, feature learning, and dimension reduction. Whereas discriminative networks capture the relationship between meta-atom geometry and spectral response from a training set, DGMs focus on learning the properties of meta-atom geometry distributions \citep{liu2018generative, hodge2019rf, jiang2019free, jiang2019global}. Major classes of DGMs (Fig.~\ref{fig:netStructures}) applied to MTS inverse design are as follows.
	
	\subsubsection{Generative adversarial networks (GANs)}
	\label{sec:GANs}
	In a GAN system, two ANNs compete to improve each of their models: the generative network learns to create inputs indistinguishable from the training data while the discriminative network learns to identify true data from the output of the generative network. Training GANs involves jointly training a generator network and a discriminator network in a game theoretic approach to find a local Nash equilibrium. The goal of a generative model is to observe a collection of training examples and learn the underlying probability distribution that generates them. GANs are able to generate new samples from the estimated probability distribution. GANs were initially applied to generate photos, however, have been applied to many domains including speech and video generation. Very recently, GANs have been applied to generate new MTS hardware design including those not explicitly seen in the training dataset or current literature. At the end of a successful training process, GANs are able to produce realistic meta-atom designs, even for very complicated datasets and spectral responses.
	
	In, \citep{hodge2019rf}, we introduce GANs to microwave MTS design. GANs are promising for low-cost MTS design with complex frequency and polarization dependent scattering responses. In \citep{liu2018generative}, an input set of user-defined EM spectra is fed to GAN that generates candidate patterns to match the on-demand spectra with high fidelity. Here, DNNs are employed to approximate the spectra of the MTS and perform inverse design by generating meta-atom structures that yield user-defined input spectra. Once the model is trained, extensive parameter scans and trial-and-error procedures are bypassed. This conditional deep convolutional GAN (cDC-GAN) architecture uses three interconnected CNNs: generator, discriminator, and simulator. The simulator is a pretrained network that serves as a surrogate model for fast spectral performance prediction. In this study, $S$ is a five-layer CNN with three-fully connected layers at the output. 
	The conditional generator networks accepts the desired spectral response and a latent noise vector to output potential meta-atom designs. The discriminator serves to train the generator by evaluating the distance of the distributions between the geometric patterns from training data and generator. At the end of successful training, discriminator is unable to distinguish batches from generator and training set. This approach is shown to exhibit high accuracy in inverse design of meta-atoms.
    
	In \citep{hodge2019rf}, a deep convolutional GAN (DC-GAN) is employed to generate anisotropic RF meta-atom designs. Using a small set of simulated spectra, the network learned the relationship between the physical structure of meta-atoms and their reflection spectra for vertical and horizontal polarizations. The DC-GANs generated meta-atom structures that resembled design features in the training data. To speed up training, the network was fed with parametric variations of twelve published meta-atom designs to a full-wave EM simulator. Starting out with parametric variations of canonical meta-atoms scatterers, the network picked up more efficiently than it would have from training with responses of randomized pixel data.
	

	\subsubsection{Conditional variational autoencoder (cVAE)}
	As an alternative to GAN approaches, \citep{ma2019probabilistic} presents a probabilistic DGM that solves both forward and inverse problems at the same time. It is trained in an end‐to‐end manner and uses a deep convolution cVAE (cDC-VAE) architecture (Fig.~\ref{fig:netStructures}) comprising an encoder-decoder network structure. The encoder maps the meta-atom structure to a multivariate Gaussian distribution in the latent space and the conditional decoder network inputs the reflection spectra and latent variable to generate meta-atom designs (Fig.~\ref{fig:netStructures}).
	
	In \citep{ma2019probabilistic}, the RIS inverse design is modeled in a probabilistic generative manner to investigate the complex structure–performance relationship in an interpretable way and solve the one-to-many mapping issue that is intractable in deterministic models. It developed a semi-supervised learning strategy that allows the model to utilize unlabeled data in addition to labeled data in an end-to-end training. The RIS design and spectral response are encoded into a low-dimensional latent space with a predefined prior distribution, from which the latent variables are sampled. The DGM, comprising prediction, recognition, and generation models, serves as a tool to accelerate the design, characterization, and even new discovery of MTS.

	\subsubsection{Global topology optimization networks (GLOnets)}
	Recently, GANs utilized to learn structural features of topology-optimized (TO) metagratings for inverse design \citep{jiang2019free, jiang2019global}. TO is a method of optimizing a material layout or an array of pixels to maximize system performance given a set of constraints and boundary conditions. Unlike other approaches, simulation overload for TO does not increase with the number of RIS units. In \citep{jiang2019free}, free-form diffractive metagratings were designed using TO-GAN. Here, DGMs were trained from images of periodic, TO metagratings to produce efficient scattering structures with the desired performance over a broad range of frequencies and angles. The network employed $5,000$ training examples for each angle. However, the performance of the best structures was not robust and additional refinement was needed to meet the desired performance.
	In \citep{jiang2019global}, dielectric metasurfaces optimization was performed using a physics-informed cGAN. Global optimization-based generative networks (GLOnets) are able to search the design space for the global optimum design. Unlike other GAN approaches, GLOnets seek to fit a narrow-peaked function centered on the optimal solution without a training set. The GLOnet generates a distribution of meta-atoms to samples the global design space and then shifts the distribution toward a more optimal design. Training requires computing forward and adjoint EM simulations of output structures using backpropagation. In this work, GLOnets are shown to be successful and computationally efficient global TO for MTS and metagratings.

    \section{Case Studies}
    \label{sec:case_studies}
	We perform two case studies for the design of single- and multi-layer RIS based on \cite{hodge2019rf} and \cite{hodge2019joint}, respectively. The design approach in \citep{hodge2019joint} introduced the cDC-GAN-based for jointly designing several layers of tensorial RIS. It represented three RIS layers with a $64\times 64\times 3$ red-green-blue (RGB) image matrix. The advantages of the cDC-GAN are that it trains classifiers in a semi-supervised manner and generates new free-form shapes not previously shown in the literature. However, GANs can be unstable and challenging to train. We validated the designs by simulating their spectrum using a full-wave EM solver and comparing the results to the desired spectrum. In this data representation, the top layer meta-atom design is represented as the first channel, the second layer meta-atom design is represented by the second channel, and a third layer is represented by the third channel using the conventional RGB image format. 
	
	\subsection{MTS Characterization Model}
	Consider a two-dimensional (2-D) MTS lying in the x-y plane with z-axis being the direction of propagation. According to Huygens' principle, the EM fields created by arbitrary sources in an arbitrary volume $V$ are found as the fields created by equivalent surface currents on the volume surface $S$ \citep{tretyakov2015metasurfaces}. Therefore, a known incident EM source, such as a plane wave, can be transformed into a desired transmitted or reflected wave using an MTS. The MTS creates the desired aperture field distribution or phase shift by modifying the effective boundary conditions of the EM surface. 
    
    The amplitude and phases of transmitted and reflected waves from MTS are functions of surface-averaged induced electric and magnetic current densities, $\boldsymbol{J}_{e}$ and $\boldsymbol{J}_{m}$, respectively. These effective surface current densities induced on MTS are described by average tangential electric ($\boldsymbol{E}_{t\pm}$) and magnetic ($\boldsymbol{H}_{t\pm}$) fields on each side of MTS as  
\begin{align}
    \boldsymbol{J}_{e} &= \boldsymbol{n} \times (\boldsymbol{H}_{t+}-\boldsymbol{H}_{t-}),  \label{eqn:A1}\\
    \boldsymbol{J}_{m} &= -\boldsymbol{n} \times (\boldsymbol{E}_{t+}-\boldsymbol{E}_{t-}),  \label{eqn:A2}
\end{align}
where $\boldsymbol{n}$ is the unit vector normal to the MTS. Examples of passive implementations of Huygens' MTS (HMS) include reflectionless refraction, perfect anomalous reflection, and arbitrary antenna beamforming \citep{chen2018huygens}. 

The induced surface currents $\boldsymbol{J}_{e}$ and $\boldsymbol{J}_{m}$ are related to their respective average tangential fields (applied on a thin slab of polarizable particles) by spatially-varying electric surface impedance $\overline{\overline{Z}}_{\textrm{se}}$ and magnetic surface admittance $\overline{\overline{Y}}_{\textrm{sm}}$,

\begin{align}  
    \boldsymbol{E}_{t,\textrm{avg}} &= \overline{\overline{Z}}_{\textrm{se}} \cdot \boldsymbol{J}_{e}, \label{eqn:B1}\\
    \boldsymbol{H}_{t,\textrm{avg}} &= \overline{\overline{Y}}_{\textrm{sm}} \cdot \boldsymbol{J}_{m}. \label{eqn:B2}
\end{align}
Substituting (\ref{eqn:A1}) and (\ref{eqn:A2}) into (\ref{eqn:B1}) and (\ref{eqn:B2}), respectively, yield the following generalized sheet transition conditions (GTSCs) \citep{holloway2012overview,epstein2016huygens} used for describing an MTS:
\begin{align}  
    \boldsymbol{E}_{t,\textrm{avg}} &= \overline{\overline{Z}}_{\textrm{se}} \cdot [\boldsymbol{n} \times (\boldsymbol{H}_{t+}-\boldsymbol{H}_{t-})], \\
    \boldsymbol{H}_{t,\textrm{avg}} &= \overline{\overline{Y}}_{\textrm{sm}} \cdot [-\boldsymbol{n} \times (\boldsymbol{E}_{t+}-\boldsymbol{E}_{t-})]. \label{eqn:C}
\end{align}
In case of single polarization, the tensor quantities above reduce to scalars \citep{epstein2016huygens}. Specifying the desired incident fields $\boldsymbol{E}_{t-}$ and $\boldsymbol{H}_{t-}$ and desired output fields $\boldsymbol{E}_{t+}$ and $\boldsymbol{H}_{t+}$ leads to computation of the required electric impedance and magnetic admittance at each (spatial) location of the MTS. The bianisotropy is included in these boundary conditions by introducing the tensor magneto-electric coupling coefficient $\overline{\overline{K}}_{em}$ as
\citep{chen2018huygens}

\begin{align}  
    \boldsymbol{E}_{t,\textrm{avg}} &= \overline{\overline{Z}}_{\textrm{se}} \cdot \boldsymbol{J}_{e}-\overline{\overline{K}}_{\textrm{em}} \cdot [\boldsymbol{n} \times \boldsymbol{J}_{m}], \\
    \boldsymbol{H}_{t,\textrm{avg}} &= \overline{\overline{Y}}_{\textrm{sm}} \cdot \boldsymbol{J}_{m}-\overline{\overline{K}}_{\textrm{em}} \cdot [\boldsymbol{n} \times \boldsymbol{J}_{e}]. \label{eqn:Bten}
\end{align}

\begin{figure}
  \centering
  \includegraphics[width=85mm]{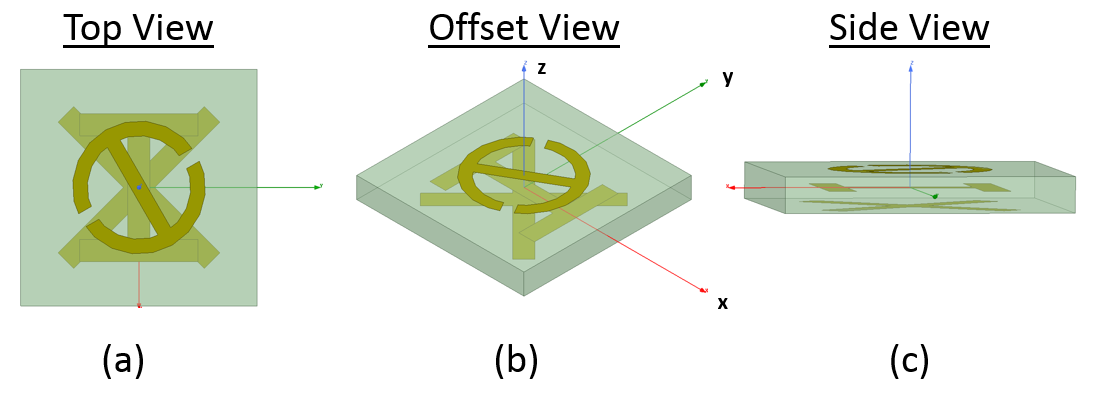}
  \caption{Illustration of a three-layer MTS unit-cell (meta-atom) in the three-dimensional x-y-z Cartesian space. The (a) top, (b) offset, and (c) side views were simulated in ANSYS HFSS software using periodic boundary conditions. 
  The meta-atom occupies a 8 mm by 8 mm planar grid. The gold color rectangles represent copper traces on the top, middle, and bottom layers of meta-atom. The dielectric spacer is made up of 1.0 mm thick duroid material with relative permittivity $\epsilon_{r}=2.2$. 
  We used Floquet ports in HFSS to excite the meta-atom unit-cell by a wave traveling in the negative z-direction  \citep{hodge2019multi}.
  }
  \label{fig:exampleHFSSunitCell}
\end{figure}
The transmission and reflection spectral responses of MTS, described by vectors $\mathbf{T}$ and $\mathbf{R}$, respectively, are functions of the surface impedances at a particular incidence angle $\theta$ and frequency $f$ (GHz). For instance, $\mathbf{T}(\theta, f) = h(\overline{\overline{Z}}_{\textrm{se}}, \overline{\overline{Y}}_{\textrm{sm}}, \overline{\overline{K}}_{\textrm{em}})$ \citep{chen2018huygens,epstein2016arbitrary} where $h(\cdot)$ is a non-linear function. In this chapter, we fix $\theta=0$ (broadside incidence) so that $\mathbf{T} =\mathbf{T}(0, f)$, where we have omitted the arguments for simplicity. From here on, we focus on only $\mathbf{T}$ because the design procedure using $\mathbf{R}$ is identical. 

The EM wave is also characterized by its polarization. We consider two polarizations - `$x$' and `$y$' - wherein the electric field is parallel to the x- and y-directions, respectively. For an incident wave with a particular polarization, the MTS produces responses in both polarizations. For example, the response in $x$ ($y$) polarization when the incident wave is also $x$-polarized ($y$-polarized) is the co-polar response $\mathbf{T}_{xx}$ ($\mathbf{T}_{yy}$). Similarly, cross-polar responses $\mathbf{T}_{xy}$ and $\mathbf{T}_{yx}$ are defined. 
A multi-layer meta-atom consists of multiple layers of different shapes separated by dielectric spacers for structural support. Consider a 3-layer MTS (Fig.~\ref{fig:exampleHFSSunitCell}) whose composite response $\mathbf{T}$ is the superposition of the responses of individual layers. Our goal is to train the MD-GAN to implicitly learn physical quantities $\overline{\overline{Z}}_{\textrm{se}}$, $\overline{\overline{Y}}_{\textrm{sm}}$, and  $\overline{\overline{K}}_{\textrm{em}}$ by mapping various design geometries to transmission spectra and produce new meta-atom designs for each layer to realize composite responses $\mathbf{T}_{xx}$, $\mathbf{T}_{yy}$, $\mathbf{T}_{xy}$, and $\mathbf{T}_{yx}$.

\subsection{Training and Design}
\begin{figure}[t]
    \centering
    \includegraphics[width=1.0\textwidth]{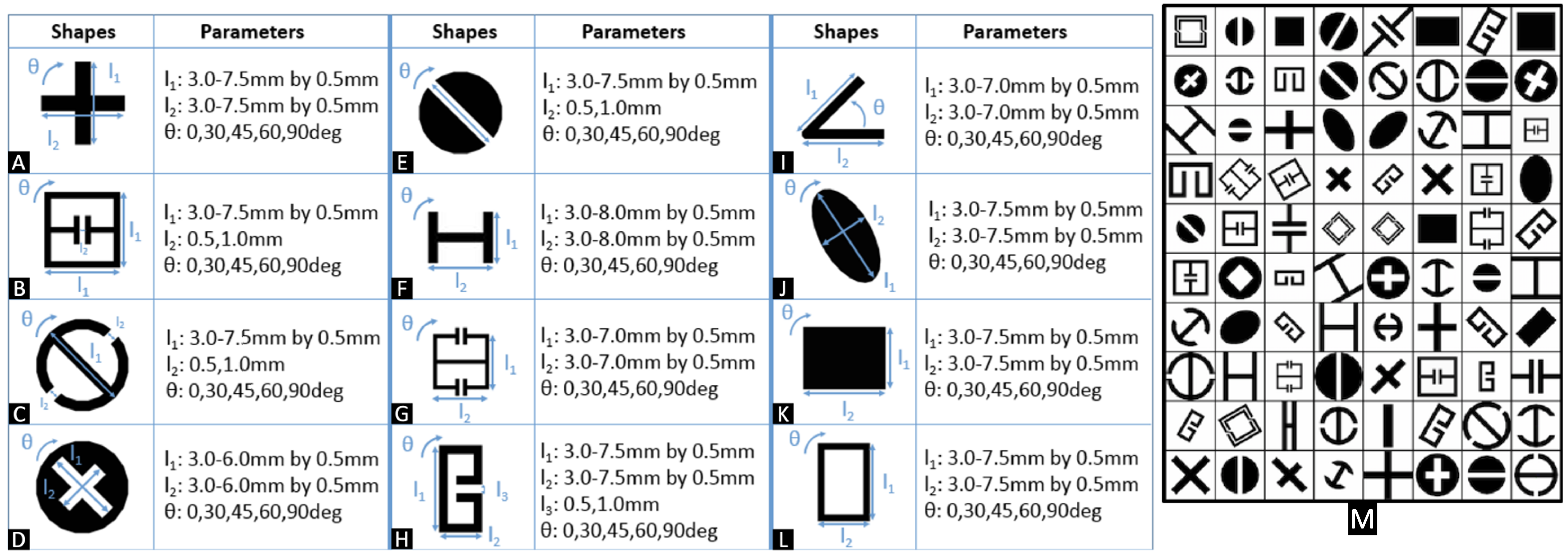} 
    \caption{\scriptsize{Meta-atom patterns from published literature. During training, several variations of these shapes are generated by changing parameter values as indicated. A. \citep{chen2012interference} B. \citep{schurig2006electric} C. \citep{su2016ultra} D. \citep{pereda2016dual} E.  \citep{minatti2015modulated} F. \citep{sun2012gradient} G. \citep{nguyen2018impedance} H. \citep{hodge2014enhancement} I. \citep{yu2011light} J. \citep{mencagli2015surface} K. \citep{fong2010scalar} L. \citep{su2017ultra}. M. 80 randomly selected training images of MTS unit cell designs from our design database used to train DC-GAN using the basic shapes A.-M. }
    \vspace{-12pt}}
    \label{fig:MetaAtomVariationDiagram_4by3_revC}
    \end{figure}

    We evaluated proposed inverse design approach by implementing our distributed cMD-GAN architecture using PyTorch and performing simulations on an NVIDIA Tesla T4 GPU. During training, we included parametric variations of only those meta-atom shapes that have been extensively studied in the literature. Fig.~\ref{fig:MetaAtomVariationDiagram_4by3_revC} lists these shapes and enumerates variations in the physical parameters to generate training data. The CNNs process matricized data e.g. a color image composed of three matrices, each of which contains pixel intensities in red, green, and blue (RGB) color channels. Rather than feeding a three-channel image matrix representing physical RGB colors as is conventionally done in image recognition, we exploit the three channel matrix input into the CNN to represent spatial design of meta-atom scatterers in different layers of a multi-layer MTS. Prior works do not employ this innovative technique of representing multiple MTS layers as channels of an image matrix.

\begin{figure}
		\centering
		\includegraphics[width=0.8\textwidth]{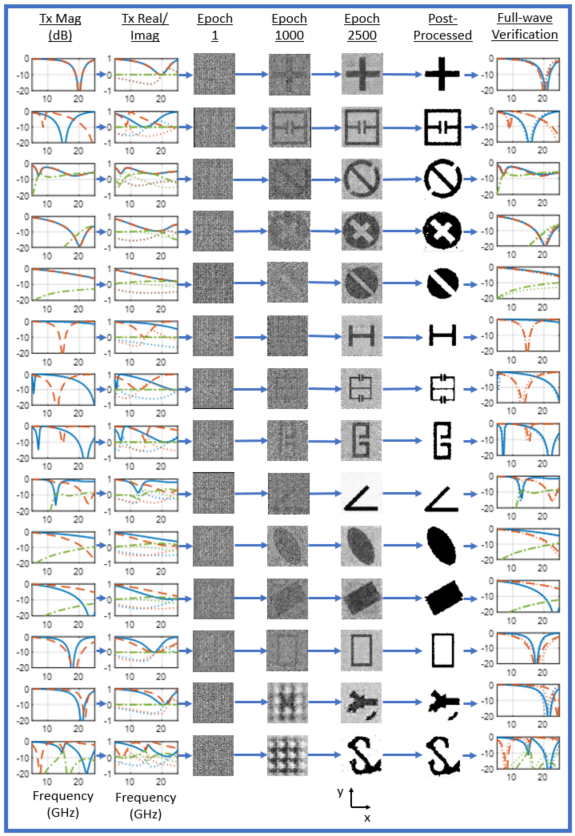}
		\caption{Meta-atom structures generated using DC-GAN \citep{hodge2019joint}. The first twelve rows show the ability of the DC-GAN to regenerate canonical structures from the training data set. The last two rows show the ability of the DC-GAN to generate new meta-atom geometries, exhibiting spatial features similar to those in the training data set \citep{hodge2021deep}.
		}
		\label{fig:numExpResults}
	\end{figure}

\begin{figure}
		\centering
		\includegraphics[width=0.5\textwidth]{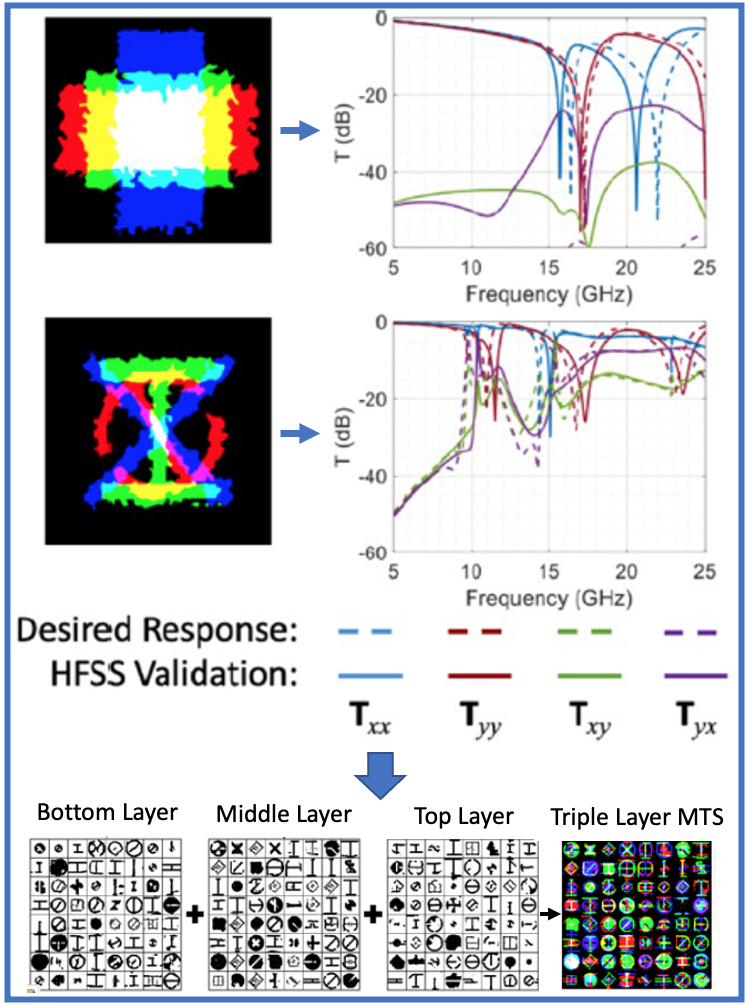}
		\caption{Two three-layer meta-atom designs (upper left) generated from cMD-GAN in \citep{hodge2019multi}. The top (red), middle (green), and bottom (blue) layers are metallic traces separated by dielectric spacers. The image matrices are post-processed to remove the background noise. The upper right shows desired input RF transmission response vectors (dashed lines) and the full-wave verification using the Ansys HFSS finite element method solver (solid lines) of generated design for $5$-$25$ GHz when illuminated by a plane wave at normal incidence. The blue and maroon lines represent the respective $x$ and $y$ co-polarized transmission responses. Similarly, the green and purple lines represent the cross-polarized responses for $x$ and $y$ polarized signals. The bottom row shows a composite of each layer of generated meta-atoms for a three-layer RIS \citep{hodge2021deep}.
		}
		\label{fig:resultsMDgan}
	\end{figure}

	In the first case study, we generated single-layer meta-atom designs using cDC-GAN. The co-polarization and cross-polarization transmission responses of the resulting meta-atom designs (Fig.~\ref{fig:numExpResults}) differed from EM simulators by less than a dB. One of the most exciting features of cDC-GAN is its ability to discover new geometries not previously found in the literature. This suggests that the model implicitly learned the physical relationships of Maxwell's equations rather than simply interpolating from past designs. We perform a second case study for multi-layer meta-atom design. Building on this techniques, the federated learning approach in \citep{hodge2019multi} employed a conditional multi-discriminator distributed GAN (cMD-GAN) (see Fig.~\ref{fig:netStructures}) for multi-layer RF MTS discovery (Fig.~\ref{fig:resultsMDgan}). The results show the feasibility of GAN-based approaches for meta-atom discovery.

	\section{Applications}
	\label{sec:app}
	Lately, the RIS-aided wireless systems have exploited DL to handle very challenging problems. For instance, signal detection in RIS requires development of end-to-end learning systems under the effect of channel and beamformers~\citep{irs_DL_detection}. The channel needs to be estimated for multiple communication links, i.e., BS-user and BS-RIS-user~\citep{elbir_LIS}. Finally, beamformers are designed (by solving complex optimization problems) for phase shifters at both BS and passive elements of the RIS~\citep{lis_channelEst_reflectedBFDesign}. The DL-based techniques are able to handle the multidimensional, huge datasets in all these problems and may also be employed for channel modeling~\citep{irs_survey1}, where the conventional model-based approaches are not very useful. There have been recent surveys on applying DL~\citep{dl_WCM} and RIS~\citep{irs_survey1} individually to wireless communications. Here, we provide an overview of systems which jointly employ both approaches. In particular, we describe DL techniques (Table~\ref{tableSummary}) for three important RIS problems: signal detection, channel estimation, and beamforming. Each of these requires  different DL architectures, which have so far included supervised learning (SL), unsupervised learning (UL), reinforcement learning (RL) and federated learning (FL). The UL and RL do not require labeling; SL needs labeled dataset; and FL has distributed structure for model training. We provide a detailed synopsis of the advantages and shortcomings of each algorithm for these three applications in the subsequent sections.

	
	\begin{figure}
		\centering
		{\includegraphics[width=1.0\textwidth]{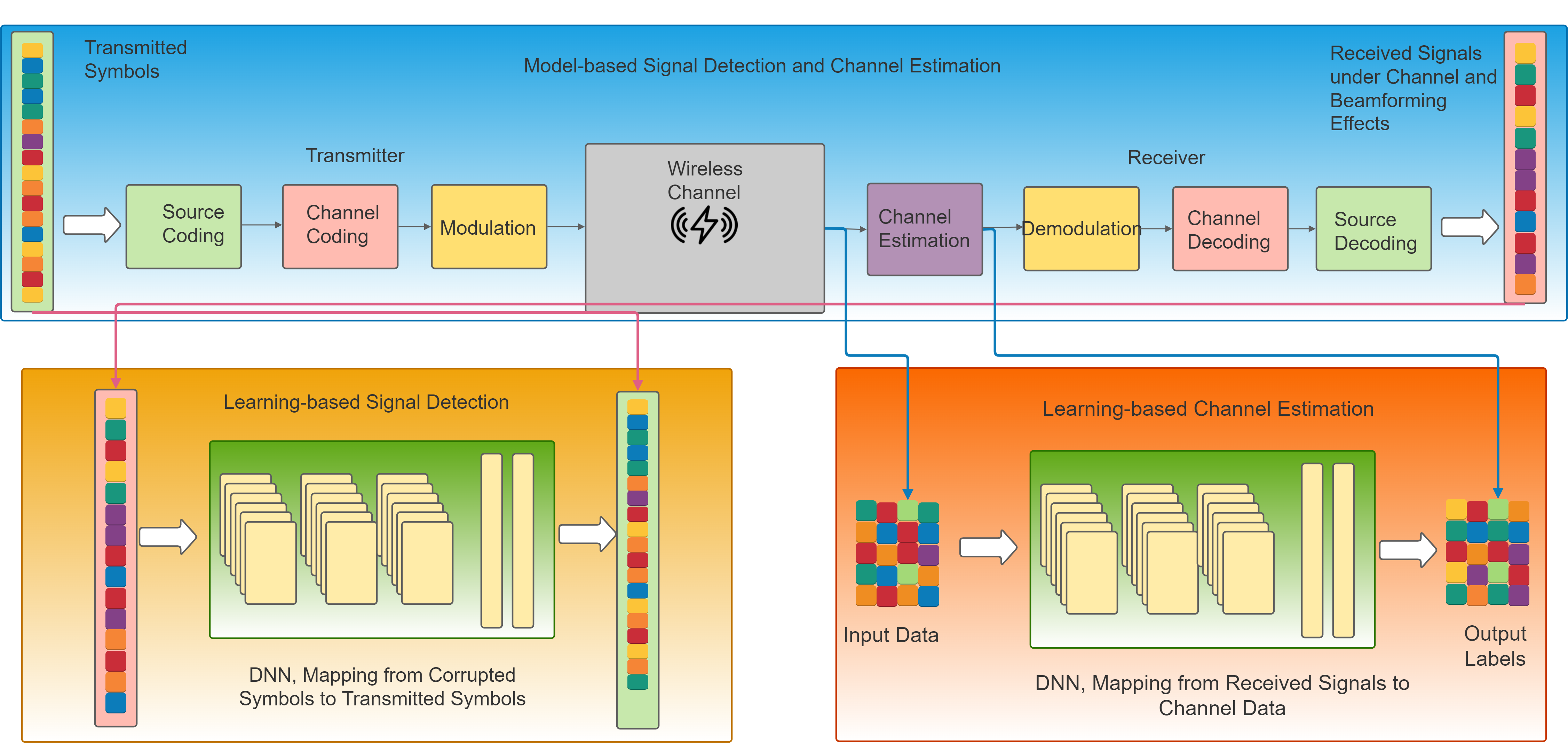}}
		\caption{Model-based versus learning-based frameworks for signal detection and channel estimation. Model-based approach (top) comprises multiple subsystems to process the received signal. Learning-based signal detection (bottom, left) provides an end-to-end data mapping from the corrupted symbols under the channel effects at the receiver to the transmit symbols. Learning-based channel estimation (bottom, right) maps the input received signals to the channel estimate as output labels \citep{elbir2020survey}. 
		}	
		\label{fig_SD_CE}
	\end{figure}

	


	%
	%
	%
	%
	%


	In RIS-assisted scenario, wherein the BS with $M$ antennas  transmits $K$ data symbols $s_k\in\mathbb{C}$ by using a baseband precoder $\mathbf{F}= [\mathbf{f}_1,\dots,\mathbf{f}_K]\in\mathbb{C}^{M\times K}$ . Hence, the downlink $M\times 1$ transmitted  signal becomes $	{\mathbf{s}} = \sum_{k=1}^{K}  {\mathbf{f}}_ks_k$. The transmitted signal is received from the $k$-user with two components, one of which is through the direct path from the BS and the another one is through the RIS. The received signal from the $k$-th user can be given by
	\begin{align}
	\label{irsSignal}
	y_k = \big(\mathbf{h}_{\mathrm{D},k}^\textsf{H} + \mathbf{h}_{\mathrm{A},k}^\textsf{H} \boldsymbol{\Psi}^\textsf{H} \mathbf{H}^\textsf{H}  \big) {\mathbf{s}} + n_k,
	\end{align}
	where $n_k\sim \mathcal{CN}(o,\sigma_n^2)$ and $\mathbf{h}_{\mathrm{D},k}\in\mathbb{C}^{M}$ denotes the direct channel between the BS and the $k$-th user. The vector $\mathbf{h}_{\mathrm{A},k}\in \mathbb{C}^L$ expresses the RIS-assisted channel between the RIS and the $k$-th user.  {\color{black}$\boldsymbol{\Psi}\in\mathbb{C}^{L\times L}$ is a diagonal matrix, i.e., $\boldsymbol{\Psi} = \mathrm{diag}\{\beta_1\exp(j \phi_1),\dots,\beta_L\exp(j \phi_L) \}$. Here, $\beta_l \in \{0,1\}$ represents the on/off state of the RIS elements. In practice, the RIS elements cannot be  perfectly turned on/off, Hence, $\beta_l$ can be modeled as $\beta_l=\left\{\begin{array}{cc}
		1 - \epsilon_1 & \mathrm{ON}\\
		0 + \epsilon_0 & \mathrm{OFF}
		\end{array}\right.$ for $\epsilon_1,\epsilon_0 \geq0$. $\phi_l\in [0,2\pi)$ is the phase shift of the reflective elements.} Finally, the channel between the RIS and the BS is represented by $\mathbf{H}\in \mathbb{C}^{M\times L}$.

	In mm-Wave transmission, the channel can be represented by the Saleh-Valenzuela (SV) model where a geometric channel model is adopted with limited scattering. Hence, we assume that the mm-Wave channels, i.e., $\mathbf{h}_{\mathrm{D},k}, \mathbf{h}_{\mathrm{A},k}$ and $\mathbf{H}$, include the contributions of $N_\mathrm{D}$, $N_\mathrm{A}$ and $N_\mathrm{H}$ paths, respectively. Thus, we can represent the channels $\mathbf{h}_{\mathrm{D},k}$ and $\mathbf{h}_{\mathrm{A},k}$ as
	$\mathbf{h}_{\mathrm{D},k} = \sqrt{\frac{M }{N_\mathrm{D}  }} \sum_{n_\mathrm{D}=1}^{ N_\mathrm{D}} \alpha_{\mathrm{D},k}^{(n_\mathrm{D})} \mathbf{a}_\mathrm{D}( \theta_{\mathrm{D},k}^{(n_\mathrm{D})}),$ and $\mathbf{h}_{\mathrm{A},k} = \sqrt{\frac{L }{N_\mathrm{A}  }} \sum_{n_\mathrm{A}=1}^{ N_\mathrm{A}} \alpha_{\mathrm{A},k}^{(n_\mathrm{A})} \mathbf{a}_\mathrm{A}( \theta_{\mathrm{A},k}^{(n_\mathrm{A})}),$
	where $\{\alpha_{\mathrm{D},k}^{(n_\mathrm{D})},\alpha_{\mathrm{A},k}^{(n_\mathrm{A})}\}$ and $\{\theta_{\mathrm{D},k}^{(n_\mathrm{D})},\theta_{\mathrm{A},k}^{(n_\mathrm{A})}  \}$ are the complex channel gains and received path angles for the corresponding channels, respectively. $\mathbf{a}_\mathrm{D}( \theta)$ and $\mathbf{a}_\mathrm{A}( \theta)$ are $M\times 1$ and $L\times 1$ steering vectors of the path angles as $\mathbf{a}_\mathrm{D}( \theta) = \frac{1}{\sqrt{M}}[e^{j \omega_0},\dots, e^{j \omega_{M-1}}]^\textsf{T}$, $\mathbf{a}_\mathrm{A}( \theta) = \frac{1}{\sqrt{L}}[e^{j\omega_0},\dots,e^{j \omega_{L-1}}]^\textsf{T}$ where $\omega_{n} = n\frac{2\pi d}{\lambda}\pi\sin(\theta)$  and $d = \lambda/2$ is the array spacing for the wavelength $\lambda$. Further, the mm-Wave channel between the BS and the RIS is given by
	\begin{align}
	\label{eq:ChannelModel}
	\mathbf{H} =  \sqrt{\frac{ M L } {N_\mathrm{H} }}\sum_{n_\mathrm{H}=1}^{N_\mathrm{H}}    \alpha^{(n_\mathrm{H})} \mathbf{a}_\mathrm{BS}( \theta_{\mathrm{BS}}^{(n_\mathrm{H})}) \mathbf{a}_\mathrm{RIS}^\textsf{H}( \theta_{\mathrm{RIS}}^{(n_\mathrm{H})}),
	\end{align}  
	where $\alpha^{(n_\mathrm{H})}\in \mathbb{C}$ denotes the complex gain and $\{\theta_{\mathrm{BS}}^{(n_\mathrm{H})},\theta_{\mathrm{RIS}}^{(n_\mathrm{H})}\}$ are the angle-of-departure (AOD) and angle-of-arrival (AOA) angles of the paths, respectively. $\mathbf{a}_\mathrm{BS}( \theta)\in\mathbb{C}^{M}$ and $\mathbf{a}_\mathrm{RIS}( \theta)\in\mathbb{C}^L$ are the steering vectors.	Let $\mathbf{G}_{k}\in \mathbb{C}^{M\times L}$ be the cascaded channel matrix between the BS and the $k$-th user as $\mathbf{G}_k = \mathbf{H}\boldsymbol{\Gamma}_k$ where $\boldsymbol{\Gamma}_k = \mathrm{diag}\{ \mathbf{h}_{\mathrm{A},k}\}$. Then, we can write $\mathbf{H} \boldsymbol{\Psi}\mathbf{h}_{\mathrm{A},k} = \mathbf{G}_k \boldsymbol{\psi}$, for which we have $\boldsymbol{\Psi} = \mathrm{diag}\{\boldsymbol{\psi} \}$. 

	\subsection{DL-Based Signal Detection in RIS}
	The signal detection comprises mapping the received symbols under the effect of channel and beamformers to transmit symbols (Fig.~\ref{fig_SD_CE}). The signal detection problem can be formulated as
	\begin{align}
	    \hat{{\mathbf{s}}} = \argmin_{{\mathbf{s}}} \|y_k - \big(\mathbf{h}_{\mathrm{D},k}^\textsf{H} + \mathbf{h}_{\mathrm{A},k}^\textsf{H} \boldsymbol{\Psi}^\textsf{H} \mathbf{H}^\textsf{H}  \big) {\mathbf{s}} \|^2,
	\end{align}
	which requires the knowledge of the channel, i.e, $\mathbf{h}_{\mathrm{D},k}, \mathbf{h}_{\mathrm{A},k}$ and $ \mathbf{H}  $. Instead, DL-based model accepts the input data $\mathbf{y}_k = [y_{k,1},\cdots,y_{k,P_T}]^\textsf{T}$, where $P_T$ is the number of collected observations. Then, the DL model is trained to construct a non-linear mapping between the corrupted data $\mathbf{y}_k$ and the clean symbols  ${\mathbf{s}}$.

	To leverage DL for signal detection, \citep{irs_DL_detection} devised a multi-layer perceptron (MLP) for mapping the channel and reflecting beamformer effected data symbols to the transmit symbols. The MLP is a feedforward neural network (NN) composed of  multiple hidden layers. The framework in \citep{irs_DL_detection} uses three fully connected layers. Once the MLP is trained on a dataset composed of received-transmitted data symbols, each user feeds the learning model with the block of received symbols. These blocks account for the effect of channel and beamformers. Then, MLP yields the estimated transmit symbols. 
	
	A major advantage of this approach is its simplicity that the learning model estimates the data symbols directly, without a prior stage for channel estimation. Thus, this method is helpful reducing the cost of channel acquisition. In~\citep{irs_DL_detection}, a bit-error-rate (BER) analysis has shown that the DL-based RIS signal detection (DeepRIS) provides better BER than the minimum mean-squared-error (MMSE) and close performance to the maximum likelihood estimator.
	
	However, a few challenges remain to achieve a reliable performance. The training data should be collected under several channel conditions and different beamformer configurations so that the trained model learns the environment well and reflects the accurate performance in different scenarios. This is a particularly challenging task because it requires collection of the training data for different user locations. As a result, DL-based signal detection demands huge training dataset collected at different channel conditions.

	\subsection{DL-Based RIS Channel Estimation}
	The RIS is composed of a huge number of reflecting elements and, therefore, channel state acquisition is a major task in RIS-assisted wireless systems. A common approach is to turn on and off each individual RIS element one-by-one while also using orthogonal pilot signals to estimate the channel between the BS and the users through RIS. In particular, RIS channel estimation via DL involves constructing a mapping between the received input signals at the user and the channel information of direct and cascaded links (Fig.~\ref{fig_SD_CE}). In this way, DL-based techniques reduce the pilot percentage and complexity in channel estimation stage~\citep{elbir2020_FL_CE}.
	

	The SL approach proposed in \citep{elbir_LIS} estimates both direct and cascaded channels via twin convolutional neural networks (CNNs). First, the received pilot signals at the user are collected by sequentially turning on the individual RIS elements. Then, the collected data are used to find the least squares estimate of the cascaded and the direct channels. Both CNNs are trained to map the least squares (LS) channel estimates to the true channel data. The upshot is that each user estimates its own channels only once and feeds the received pilot data (LS estimate) to the trained CNN models. The CNNs have higher tolerance than MLP against the channel data uncertainties, imperfections (such as switching mismatch) of RIS elements.
	
	When the model training is conducted at the user with huge datasets as in~\citep{elbir_LIS}, the system may lack sufficient computational capability. This is overcome by FL-based training \citep{elbir2020_FL_CE}, where the learning model updates are computed at the devices (nodes) and aggregated at the BS (central server) (Fig.~3), thereby eliminating the transmission of raw data.   FL significantly reduces the transmission overhead since the size of the datasets is usually larger than the size of the learning model, and its performance improves as the number of users increases~\citep{elbir2020_FL_CE,irs_FL_BF_fromChannel}. Furthermore, instead of using two CNNs as in~\citep{elbir_LIS}, a single CNN in~\citep{elbir2020_FL_CE} jointly estimates both cascaded and direct channels.

	Although FL reduces the transmission overhead during model training, its training performance is upper bounded by the centralized model training, i.e., training the model with the whole dataset at once. Therefore, the prediction performance of FL is usually poorer than the centralized learning (CL). As shown in Fig.~\ref{fig_NMSE_ALL}), CL and FL frameworks are compared with the MMSE and the LS estimation. We note that FL performs slightly poorer than CL in high SNR regimes. Despite this, FL significantly reduces the transmission overhead, e.g., approximately ten-fold reduction in the number transmitted symbols~\citep{elbir2020_FL_CE}. The performance of FL improves with the increase in the number of users or edge devices because this reduces the variance of the model updates aggregated at the BS. The diversity of the local dataset of the users also affects the training/prediction performance and better performance is obtained if the local datasets are close to uniformity.

	Both SL- and FL-based channel estimation techniques suffer from high channel training overhead. In this context, compressive channel estimation with deep denoising neural networks (DDNNs) is very effective \citep{irs_deepDenoisingNN_CE}. It employs a hybrid passive/active RIS architecture, where the active RIS elements are used for uplink pilot training and passive ones for reflecting the signal from the BS to the users. Once the BS collects the compressed received pilot measurements, complete channel matrix is recovered through sparse reconstruction algorithms such as orthogonal matching pursuit (OMP). Then, DDNN is used to improve the channel estimation accuracy by exploiting the correlation between the real and imaginary parts of the mm-Wave channel in angular-delay domain. During training, the input is the OMP-reconstructed channel matrix and the output is the noise, i.e., the difference between the OMP estimate and the ground truth channel data. 
	This method leverages both CS and DL yielding a performance better than using these techniques individually. The major drawback is the additional hardware complexity introduced by the active RIS elements. 
	Furthermore, OMP algorithm is used in place of the raw received pilot measurements for constructing the input. This requires repeated execution of the OMP algorithm thereby increasing the prediction complexity over the DL methods in~\citep{elbir_LIS} and~\citep{elbir2020_FL_CE}.

	Consider the downlink scenario where the BS transmits the orthogonal pilot signals $\mathbf{x}_p\in\mathbb{C}^M$, one at a single coherence time $\tau$, with $p= 1,\dots,P$ and $P\geq M$.  Hence, the total number of  channel uses to estimate the direct channel is $P$. The received signal at the $k$-th user can be given by
	\begin{align}
	\label{receivedDirectChannel}
	\mathbf{y}_{k} = \big(\mathbf{h}_{\mathrm{D},k}^\textsf{H} + \boldsymbol{\psi}^\textsf{H}\mathbf{G}_k^\textsf{H}  \big)\mathbf{X} + \mathbf{n}_{k},
	\end{align}
	where $\mathbf{X} = [\mathbf{x}_1,\dots, \mathbf{x}_P]\in\mathbb{C}^{M\times P}$ is the pilot signal matrix while $\mathbf{y}_k = [{y}_{k,1},\dots, {y}_{k,P}]$ and $\mathbf{n}_k = [{n}_{k,1},\dots, {n}_{k,P}]$ are ${1\times P}$ row vectors and  $\mathbf{n}_k\sim \mathcal{CN}(0,\sigma_n^2\boldsymbol{\mathrm{I}}_{{P}})$.  We assume that the pilot training has two phases: direct channel estimation (i.e., $\mathbf{h}_{\mathrm{D},k}$) and the cascaded channel estimation (i.e., $\mathbf{G}_k$). In phase I, we assume that all of the RIS elements are turned off, i.e.,  $\beta_l = 0, \forall l$, by using the BS backhaul link. {\color{black}We note here that by setting $\beta_l$ as $\{1,0\}$ does not affect the the direct and cascaded channels since they do not depend on the reflect beamformer $\boldsymbol{\Psi}$ as seen in (\ref{receivedDirectChannel}).} Then, the received baseband signal at the $k$-th user becomes
	\begin{align}
	\label{receivedPilot_DC}
	\mathbf{y}_{\mathrm{D}}^{(k)} =\mathbf{h}_{\mathrm{D},k}^\textsf{H} \mathbf{X} + \mathbf{n}_{\mathrm{D},k}.
	\end{align}
	Here, the direct channel $\mathbf{h}_{\mathrm{D},k}$ is selected as the label of the deep network with the corresponding input data of  $	\mathbf{y}_{\mathrm{D}}^{(k)}$. 
	
	Once ${\mathbf{h}}_{\mathrm{D},k}$, being the estimated channel, is obtained, in the second phase of the training stage, the cascaded channel $\mathbf{G}_k$ can be estimated. {\color{black}This can be achieved via two approaches. In the first approach, $P=M$ pilot signals are transmitted} when each of the RIS elements is turned on one by one.  In this case, the BS sends a request to RIS via the micro-controller device in the backhaul link to turn on a single RIS element at a time.	For the $l$-th frame, the reflect beamforming vector becomes $\boldsymbol{\psi}^{(l)} = [0,\dots, 0, \psi_l, 0, \dots, 0]^\textsf{T}$ where {\color{black}$\beta_{\bar{l}} =\{ 0:\bar{l}=1,\dots,L, \bar{l} \neq l\}$} and the received signal from the cascaded channel at the $k$-th user becomes
	\begin{align}
	\label{receivedPilot_CC}
	\mathbf{y}_\mathrm{C}^{(k,l)} =  \big(\mathbf{h}_{\mathrm{D},k}^\textsf{H} + \mathbf{g}_{k,l}^\textsf{H}  \big) \mathbf{X} + \mathbf{n}_{k,l},
	\end{align}
	where $\mathbf{y}_\mathrm{C}^{(k,l)}  =[y_{\mathrm{C},1}^{(k,l)},\dots, y_{\mathrm{C},P}^{(k,l)}]$ and $\mathbf{n}_{k,l} =[n_{k,1}^{(l)},\dots, n_{k,P}^{(l)}]$ are $1\times P$ row vectors. In (\ref{receivedPilot_CC}), $\mathbf{g}_{k,l}$ represents the $l$-th column of $\mathbf{G}_k$ as $\mathbf{g}_{k,l}=\mathbf{G}_{k}\boldsymbol{\psi}^{(l)}$. 
	Then the least-squares (LS) estimate of $\mathbf{g}_{k,l}$ becomes
	\begin{align}
	\label{cascadedChannelEst}
	\hat{\mathbf{g}}_{k,l} =  \big(\mathbf{y}_\mathrm{C}^{(k,l)}\mathbf{X}^\textsf{H} \big(\mathbf{X} \mathbf{X}^\textsf{H} \big)^{-1}\big)^\textsf{H}   -  \mathbf{h}_{\mathrm{D},k}.
	\end{align}
	By using $\hat{\mathbf{h}}_{\mathrm{D},k}$, (\ref{cascadedChannelEst}) can be solved for $l = 1\dots,L$. Then, we can construct the estimated cascaded matrix as $\hat{\mathbf{G}}_k = [\hat{\mathbf{g}}_{k,1},\dots, \hat{\mathbf{g}}_{k,L}]$.
	
	Then, the deep network accepts the received signals as input  at the preamble stage. As a result, the input-output pairs become $\{\mathbf{y}_{\mathrm{D}}^{(k)}, \mathbf{h}_{\mathrm{D},k}\}$ and $\{\mathbf{y}_\mathrm{C}^{(k,l)},\mathbf{g}_{k,l}\}$ for direct and cascaded channel estimation, respectively.
	
	Now, let us consider model training via CL for channel estimation, wherein the training is performed by collecting the local datasets $\{\mathcal{D}_k\}_{k\in \mathcal{K}}$ from the users. Once the BS has collected the whole dataset $\mathcal{D} = \{\mathcal{D}_k\}_{k\in \mathcal{K}}$, the training is performed by solving the following problem {\color{black}
		\begin{align}
		\label{lossML}
		\minimize_{\boldsymbol{\theta}}  & \hspace{10pt} 
		\mathcal{L}(\boldsymbol{\theta}) \nonumber\\
		\subjectto & \hspace{10pt}f(\mathcal{X}^{(i)}|\boldsymbol{\theta}) = \mathcal{Y}^{(i)}, i = 1,\dots, \textsf{D},
		\end{align}
		where  $\textsf{D} = |\mathcal{D}|$ is the number of training samples and $\mathcal{L}(\boldsymbol{\theta})$ denotes the loss function defined  as
		\begin{align}
		\mathcal{L}(\boldsymbol{\theta}) =  \frac{1}{\textsf{D}}\sum_{i=1}^\textsf{D}\| f( \mathcal{X}^{(i)}|\boldsymbol{\theta}) - \mathcal{Y}^{(i)}  \|_\mathcal{F}^2,
		\end{align}
		which is the MSE between the label data $\mathcal{Y}^{(i)}$ and the prediction of the CNN, $f( \mathcal{X}^{(i)}|\boldsymbol{\theta})$.

		On the other hand, in FL, the local datasets $\mathcal{D}_{k\in \mathcal{K}}$ are preserved at the users and not transmitted to the BS. Hence, FL-based model training is performed at the user side as
		\begin{align}
		\label{lossFL}
		\minimize_{\boldsymbol{\theta}}  & \hspace{10pt} 
		\bar{\mathcal{L}}(\boldsymbol{\theta})  = \frac{1}{K}\sum_{k=1}^{K} \mathcal{L}_k(\boldsymbol{\theta})  \nonumber\\
		\subjectto & \hspace{10pt}f(\mathcal{X}_k^{(i)}|\boldsymbol{\theta}) = \mathcal{Y}_k^{(i)}, i = 1,\dots, \textsf{D}_k, k\in \mathcal{K},
		\end{align}
		where $\mathcal{L}_k(\boldsymbol{\theta}) =  \frac{1}{\textsf{D}_k}\sum_{i=1}^{\textsf{D}_k}\| f( \mathcal{X}_k^{(i)}|\boldsymbol{\theta}) - \mathcal{Y}_k^{(i)}  \|_\mathcal{F}^2$. Notice that the FL-based model training in (\ref{lossFL}) is solved at the user while the CL problem in (\ref{lossML}) is handled at the BS.}  To efficiently solve (\ref{lossFL}) and (\ref{lossML}), gradient descent (GD) is employed and the problems are solved iteratively. In CL, the gradient is computed over the whole dataset as $	 \mathbf{g}(\boldsymbol{\theta}_t) = \nabla \mathcal{L}(\boldsymbol{\theta}_t)$	and the parameter update is performed as 
	\begin{align}
	\label{eq:UpdateCL}
	\boldsymbol{\theta}_{t+1} = \boldsymbol{\theta}_t - \eta  {\mathbf{g}}(\boldsymbol{\theta}_t),
	\end{align}
	where $\eta$ is the learning rate.
	
	In FL, each user computes the gradients individually as $	\mathbf{g}_k(\boldsymbol{\theta}_t) = \nabla \mathcal{L}_k(\boldsymbol{\theta}_t)$ to solve (\ref{lossFL}), then sends them to the BS, where the model parameters are updated as
	\begin{align}
	\label{eq:UpdateAtBSNoiseFree}
	\boldsymbol{\theta}_{t+1} = \boldsymbol{\theta}_t - \eta  \frac{1}{K} \sum_{k=1}^{K} {\mathbf{g}}_k(\boldsymbol{\theta}_t).
	\end{align}
	Once the model is trained, each user can feed its received pilots signals to the CNN to predict its channel data.


\begin{table}
\centering
	\caption{DL-based techniques for RIS-assisted wireless systems \citep{elbir2020survey}
	\label{tableSummary}}
	{\tiny
	\begin{tabular}{||p{0.20\linewidth} || p{0.2\linewidth} | p{0.33\linewidth} | p{0.32\linewidth} || }
		\hline
		\textbf{\scriptsize{Learning Scheme}} &
		\textbf{\scriptsize{NN Architecture}} &
		\textbf{\scriptsize{Benefits}} &
		\textbf{\scriptsize{Drawbacks}}
		\\
		\hline 
		\multicolumn{4}{|c|}{\textbf{Signal detection}} \\ 
		\hline 
		SL~\citep{irs_DL_detection} 
		& 	 MLP with $3$ layers 
		&    No need for channel estimation algorithm
		&Still needs to design beamformers and requires huge datasets and deeper NN architectures   
		\\
		\hline
		\multicolumn{4}{|c|}{\textbf{Channel estimation}}\\
		\hline
		SL~\citep{elbir_LIS} 
		&	  Twin CNNs with  $3$ convolutional, $3$ fully connected layers
		&  Each user estimates its own channel with the trained model
		&  Data collection requires channel training by turning on/off each RIS elements
		\\
		\hline
		 
		FL \citep{elbir2020_FL_CE}
		& 	 A single CNN with $3$ convolutional, $3$ fully connected layers
		&  Less transmission overhead for training, A single CNN estimates both cascaded and direct channels
		& Performance depends on the number of users and the diversity of the local datasets 
		\\
		\hline
		SL\citep{irs_deepDenoisingNN_CE}	 
		& 	DDNN with $15$ convolutional layers
		&  Leverages both compressed sensing (CS) 
		and DL methods
		&  Requires active RIS elements. High prediction complexity arising from CS algorithms 
		\\
		\hline
		\multicolumn{4}{|c|}{\textbf{Beamforming}}\\
		\hline
		 
		SL \citep{lis_channelEst_reflectedBFDesign}
		& 	MLP with $4$ layers 
		&    Reduced pilot training overhead
		&  Requires active RIS elements for channel training
		\\
		\hline
		 
		UL \citep{irs_UnsupervisedLearning}
		& 	 MLP with $5$ layers
		&   Reduced complexity at the model training stage
		&  Implicitly needs the reflect beamformers as labels 
		\\
		\hline
		 
		RL \citep{irs_BF_RL_standalone_alkhateeeb} 
		& 	 DQN with $4$ layers
		&   Provides standalone operation since RL does not require labels like SL
		&   Longer training. Active RIS elements needed for channel acquisition
		\\
		\hline
		RL \citep{irs_RL_BF_IRSonly}
		&  DDPG	with $4$-layered actor and critic networks
		&   Better performance than DQN 
		& Large number of NN parameters are involved 
		\\
		\hline
		RL \citep{irs_DL_mag1} 
		&  DDPG with actor and critic networks
		&  Accelerated learning performance with the aid of optimization, shrinking the search space
		&  Additional optimization tools needed
		\\
		\hline
		FL \citep{irs_FL_BF_fromChannel} 
		&	  MLP with $6$ layers
		&   Less transmission overhead 
		involved during model training
		&  RIS must be connected to the PS
		\\
		\hline
		\multicolumn{4}{|c|}{Secure beamforming}\\
		\hline
		RL \citep{irs_BF_RL_jointBF_BS_IRS} 
		& 	DQN with $3$ layers
		& Robust against eavesdropping
		&  High model training complexity
		\\
		\hline
		\multicolumn{4}{|c|}{\textbf{Energy-efficient beamforming}}\\
		\hline
		RL \citep{irs_RL_energyEfficient_} 
		& 	 DQN
		& Energy-efficient and robust against channel uncertainty
		&  RIS beamforming only  
		\\
		\hline
		\multicolumn{4}{|c|}{\textbf{Indoor beamforming}}\\
		\hline
		SL \citep{irs_Indoor_locationToBF} 
		& MLP with $5$ layers
		& Reduces hardware complexity of multiple BSs and improves RSS 
		for indoor environments
		&  Learning model performance relies on room conditions
		\\
		\hline
		\hline 
	\end{tabular}
	}{}	
\end{table}\normalsize

	\section{DL-Aided Beamforming for RIS Applications}
	\label{sec:beamforming}
	Beamforming in RIS-based communications has diverse applications such as RIS-only beamforming (passive), BS-RIS beamforming (active/passive), secure beamforming (eavesdroppers included), energy-efficient beamforming, and indoor RIS beamforming. There are specific DL challenges and solutions to each one of these problems. 
	
	In general, the beamforming design problem in RIS-assisted scenario maximizes the spectral efficiency of the system as
	\begin{align}
	    &\maximize_{\mathbf{f}_k,\forall k} \sum_{k=1}^K \log_2( 1 + \textsf{SINR}_k) \nonumber\\
	    & \subjectto |\beta_l | =1, \phi_l \in \mathcal{S}_{\phi} \nonumber \\
	    &\hspace{30pt} \sum_{k=1}^K \|\mathbf{f}_k \|^2 \leq P_t,
	\end{align}
	where $P_t$ is the total transmit power and $\mathcal{S}_{\phi}$ denotes the set of discrete phase-shifts.  Also, we define the signal-to-interference-plus-noise ratio (SINR) as $\textsf{SINR}_k = \frac{\mathbf{h}_{\mathrm{A},k}^\textsf{H} \boldsymbol{\Psi}\mathbf{f}_k s_k }{ \sum_{n=1, n\neq k}^K  \mathbf{h}_{\mathrm{A},k}^\textsf{H} \boldsymbol{\Psi}\mathbf{f}_n s_n  + \sigma_n^2 } $, wherein only RIS-reflected channel is assumed. 

	\subsection{Beamforming at the RIS}
	The RIS beamforming requires passive elements continuously to reliably reflect the BS signal to the users.	Here, the MLP architecture \citep{lis_channelEst_reflectedBFDesign} is helpful in designing the reflect beamforming weights using active RIS elements \citep{irs_deepDenoisingNN_CE}. These elements are randomly distributed through the RIS. They are used for pilot training, after which compressed channel estimation is carried out using OMP. During data collection, the reflect beamforming weights are optimized by using the estimated channel data. Finally, a training dataset is constructed with channel data and reflect beamformers as the input-output pairs for an SL framework. Note that the active RIS elements present similar shortcomings as in~\citep{irs_deepDenoisingNN_CE}. However, the method in~\citep{lis_channelEst_reflectedBFDesign} excels by leveraging DL for designing beamformers. 

	The labeling process in~\citep{lis_channelEst_reflectedBFDesign} demands solving an optimization problem for each channel instance in training data generation stage. One possible way to mitigate this is to use label-free techniques, such as UL.	The UL approach in \citep{irs_UnsupervisedLearning} for reflect beamforming design employs MLP with five fully connected layers. The network maps the vectorized cascaded and direct channel data input to the output comprising the phase values of the reflect beamformers. The loss function is selected as the negative of the norm of the channel vector, which may seem like an unsupervised approach because it does not minimize the error between the label and learning model prediction. However, this technique yields the phase information at the output uniquely for each training samples. Consequently, the beamformers implicitly behave like a label in the training process. In UL, the training data is clustered into smaller sets without a prior knowledge about the ``meaning'' of each clustered sets. However, in \citep{irs_UnsupervisedLearning}, the output of the NN is a design parameter, i.e. reflect beamformer phases, which have the complexity of beamformer optimization for each input.

	\begin{figure}
		\centering
		{\includegraphics[width=0.60\textwidth]{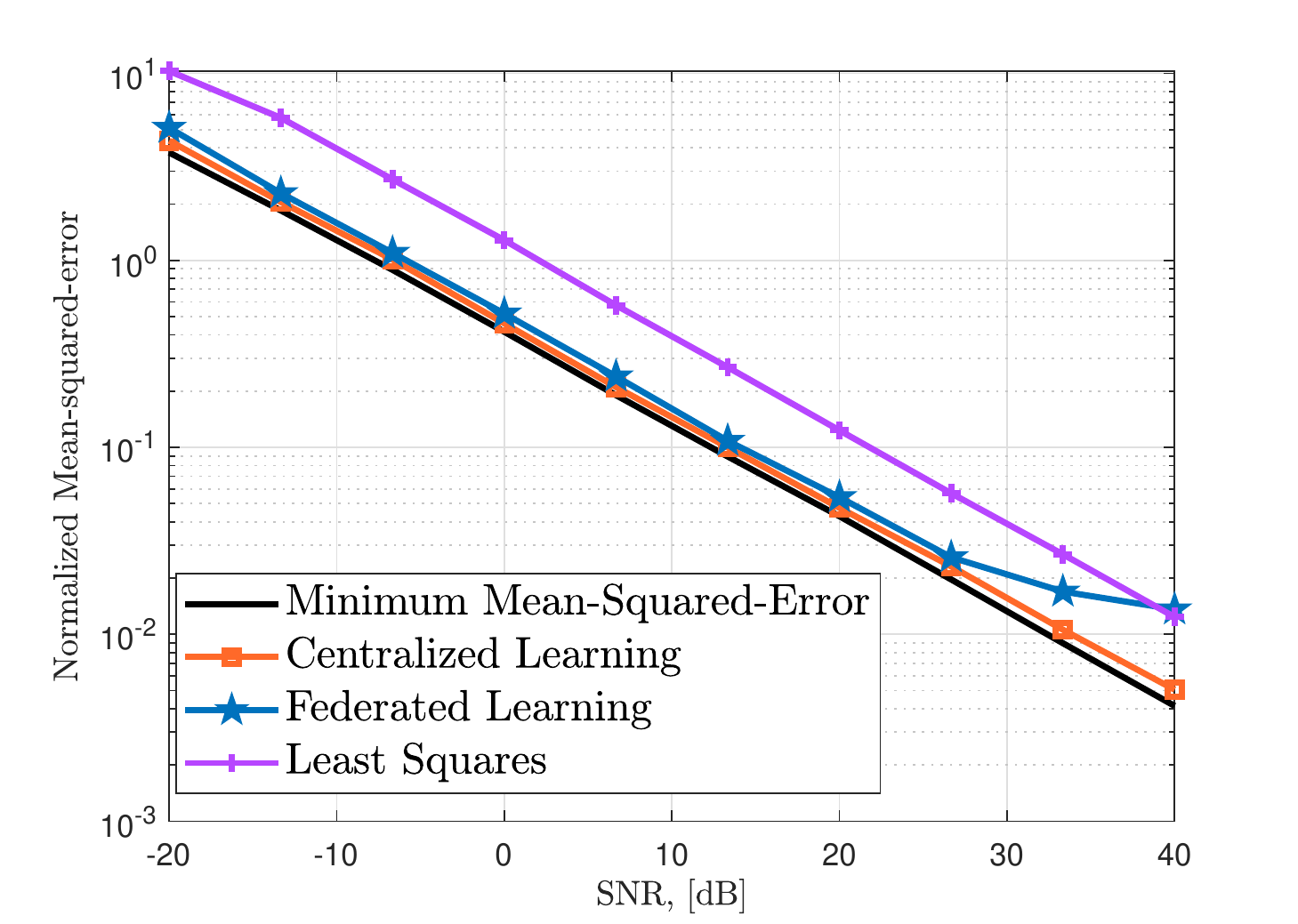} } 
		\caption{The mean-squared-error of channel estimates normalized against ground truth channel, obtained using CNN in centralized and federated learning frameworks, MMSE and LS. The BS consisted of $64$ antennas and RIS employed $64$ passive reflecting elements \citep{elbir2020survey}.
		}
		\label{fig_NMSE_ALL}
	\end{figure}

	In order to eliminate the expensive labeling process of the SL-based techniques, \citep{irs_BF_RL_standalone_alkhateeeb} employed RL to design the reflect beamformers for single-antenna users and BS. The RL is a promising approach which directly yields the output by optimizing the objective function of the learning model. First, the channel state is estimated by using two orthogonal pilot signals. An action vector is selected either by exploitation (using prior experience of the learning model) or exploration (using a predefined codebook). After computing the achievable rate based on the selected action vector from the environment, a reward or penalty is imposed by comparing with the achievable rate with a threshold. Upon reward calculation, a Deep Quality Network (DQN) (Fig.~\ref{fig_DQN_DDPG}) updates the map from the input state (channel data) to the output action (action vector composed of reflect beamformer weights). This process is repeated for several input states until the learning model converges. While RL is not an RIS-specific technique, it is particularly useful in lowering the overhead of labeling process as compared to SL architectures deployed by RNN or CNN models, which require labeled datasets.  The RL algorithm learns reflect beamformer weights based on the optimization of the achievable rate. Thus, RL presents a solution for online learning schemes, where the model effectively adapts to the changes in the propagation environment. However, RL techniques have longer training times than the SL approaches because reward mechanism and discrete action spaces make it difficult to reach the global optimum. The label-free process implies that the RL usually has slightly poorer performance than the SL.

	\begin{figure}
		\centering
		{\includegraphics[width=0.8\textwidth]{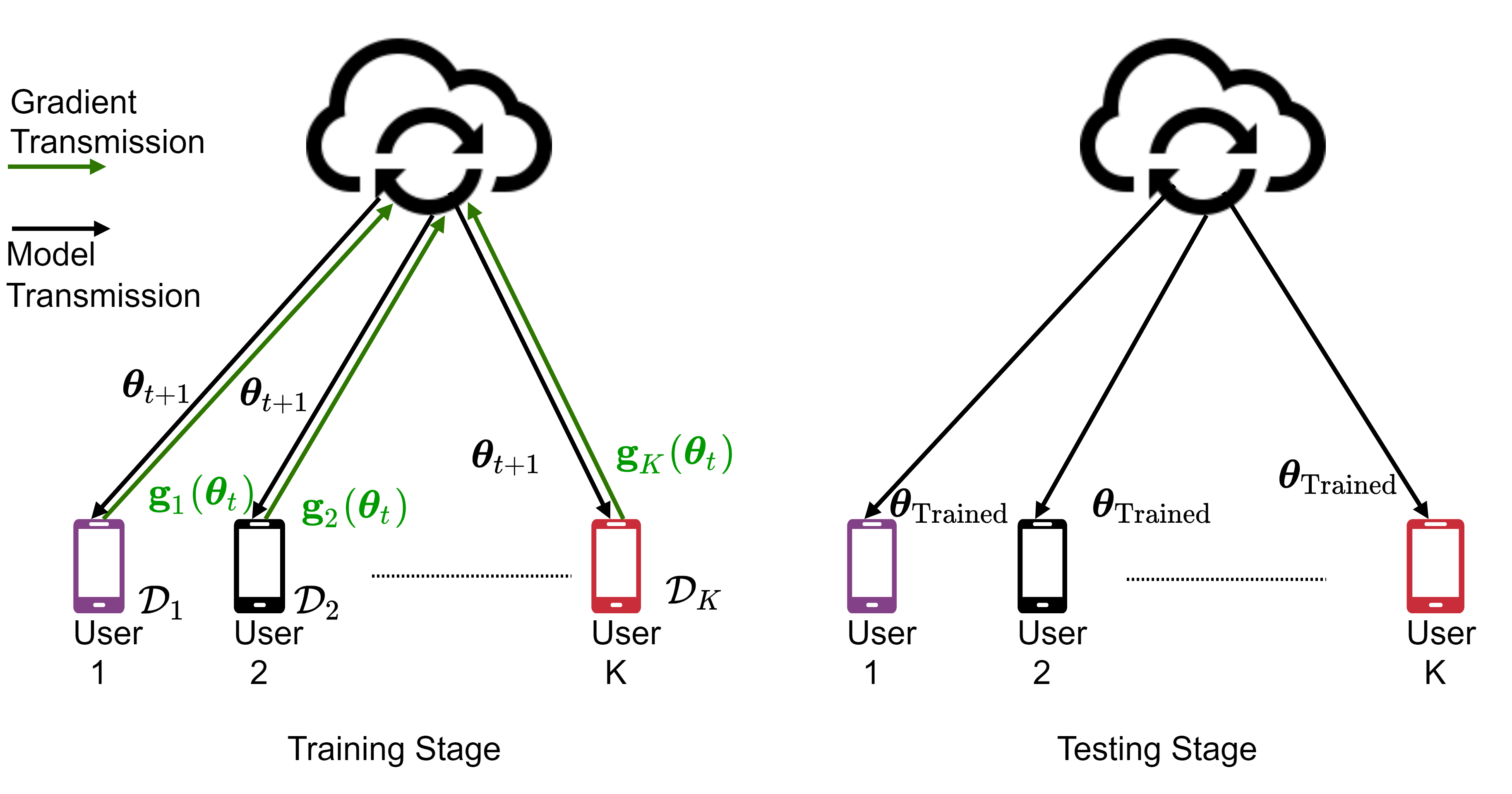} }
		\caption{In the FL framework, each user  processes its own local dataset $\mathcal{D}_k$, computes the model updates (gradients) $\mathbf{g}_k(\boldsymbol{\theta}_t)$, and sends them to the PS. The server aggregates the collected model updates and the updated model parameters are sent back to the users as $\boldsymbol{\theta}_{t+1}$\citep{elbir2020survey}.}
		\label{fig_FL}
	\end{figure}
	
	To accelerate the training stage by the use of continuous action spaces, a deep deterministic policy gradient (DDPG) (Fig.~\ref{fig_DQN_DDPG}) was introduced in \citep{irs_RL_BF_IRSonly}. Here, actor-critic network architectures are used to compute actions and target values, respectively. 	First, the learning stage is initialized by the use of input state excited by cascaded and direct channels. Given the state information, a deep policy network (DPN) (actor) constructs the actions (reflection beamformer phases). Here, the DPN provides a continuous action space that converges faster than the DQN architecture in~\citep{irs_BF_RL_standalone_alkhateeeb}. The action vector is used by the critic network architecture to estimate the received signal-to-noise-ratio (SNR) as objective. This SNR then yields the target beamformer vector under the learning policy. Using the gradient of DPN, the network parameters are updated and the next state is constructed as the combination of the received SNR and the reflecting beamformers. This process is repeated until it converges. An additional benefit of this approach is that it outperforms fixed-point iteration (FPI) algorithms used to solve reflect beamforming optimization. Moreover, the continuous action space representation with DPN in DDPG provides robustness of the learning model against changes in channel data. 
	However, multiple NN architectures (actor and critic networks) increase the number of learning parameters and aggravate model update requirements for each architecture.

	The model initialization in both DQN and DDPG may force the learning models to start far from the optimum point during the early stages of learning. This leads to a slow convergence and poor reward performance. In order to accelerate the learning process, \citep{irs_DL_mag1} devised a joint learning and optimization technique. The key idea is to use DDPG to search for optimal action for each decision epoch during training. Then, a feasible beamformer vector is found via optimization in a convex-approximation setting. This reduces the search space of the DDPG algorithm and shortens training times. 
	
	Even if RL is a label-free approach that reduces the overhead during training data generation, training approaches in~\citep{irs_RL_BF_IRSonly,irs_BF_RL_standalone_alkhateeeb,irs_DL_mag1} demand expensive transmission overhead to be trained on huge datasets. This is mitigated in FL techniques.	The FL approach in~\citep{irs_FL_BF_fromChannel} learns the RIS reflect beamformers by training an MLP by computing the model updates at each user with the local dataset. The model updates are aggregated in a parameter server (PS), which is connected to the RIS. The MLP input is the cascaded channel information and the output labels are RIS beamformer weights. The federated architecture lowers the transmission overhead during training. However, it is assumed that the PS is connected to the RIS. The simple architecture of the RIS could make this infeasible. It is more practical to access the PS via BS for model training.

	\subsection{Secure-Beamforming}
	Physical layer security in wireless systems is largely achieved through signal processing techniques, such as cooperative relaying and cooperative jamming. The hardware complexity is a major issue in these methods. The low-cost, less complex RIS-based systems have the potential to mitigate these problems. The RL-based secure beamforming \citep{irs_BF_RL_jointBF_BS_IRS} minimizes the secrecy rate by jointly designing the beamformers at the RIS and BS to serve multiple legitimate users in the presence of eavesdroppers. The RL algorithm accepts the states as the channel information of all users, secrecy rate and transmission rate. Similar to~\citep{irs_RL_BF_IRSonly}, the action vector are beamformers at the BS and RIS. The reward function is designed based on the secrecy rate of users. 
	A DQN is trained to learn the beamformers by minimizing the secrecy rate while guaranteeing the quality-of-service requirements. The model training takes place at the BS, which is responsible for collecting the environment information (channel data) and making decisions for secure beamforming. This scheme is more realistic and reliable than that of~\citep{irs_BF_RL_standalone_alkhateeeb,irs_RL_BF_IRSonly}, which ignore the effect of eavesdroppers. The learning model includes high-dimensional state and action information, such as the channels of all users and beamformers of BS and RIS. This may necessitate more computing resources for training than non-secure RIS~\citep{irs_BF_RL_standalone_alkhateeeb,irs_RL_BF_IRSonly} and conventional SL techniques~\citep{elbir_LIS,lis_channelEst_reflectedBFDesign}.
	
	\subsection{Energy-Efficient Beamforming}
	The RIS configuration dynamically changes depending on the network status. It is very demanding for the BS to optimize the transmit power every time when the on/off status of RIS elements is updated. This could be addressed by accounting energy-efficiency in the beamformer design problem. In~\citep{irs_RL_energyEfficient_}, a self-powered RIS scenario maximizes the energy-efficiency by optimizing the transmit power and the RIS beamformer phases. In this DQN-based RL approach, the BS learns the outcome of the system performance while updating the model parameters. Thus, the BS makes decisions to allocate the radio resources by relying on only the estimated channel information. The RL framework has states selected as the estimated channels from users and the energy level of the RIS. Meanwhile, the action vector includes the transmit power, the RIS beamformer phases and on/off status of the RIS elements. The learning policy is based on the reward which is selected as the energy-efficiency of the overall system. 
	However, this work considers only RIS beamforming and ignores the same at the BS.

	\subsection{Beamforming for Indoor RIS}
	Different from the above scenarios, \citep{irs_Indoor_locationToBF} addresses the RIS beamformer design problem in an indoor communications scenario to increase the received signal strength (RSS) (see Fig.~\ref{fig_IRS}). This is particularly useful from the perspective of low hardware complexity because it eliminates deployment of multiple BSs to improve RSS. The MLP architecture in \citep{irs_Indoor_locationToBF} accepts two-dimensional user position vector and yields the RIS beamformer phases at the output. Since the channel data is not employed as input, the network does not have to deal with severe environmental fluctuations.	However, the learning model trains on specific room environments and may perform poorly for different room conditions or different obstacle distribution in the same room. This is mitigated in RL-based solutions which are highly adaptive to different environments~\citep{irs_BF_RL_standalone_alkhateeeb,irs_RL_BF_IRSonly}.

	\begin{figure}
		\centering
		{\includegraphics[width=0.8\textwidth]{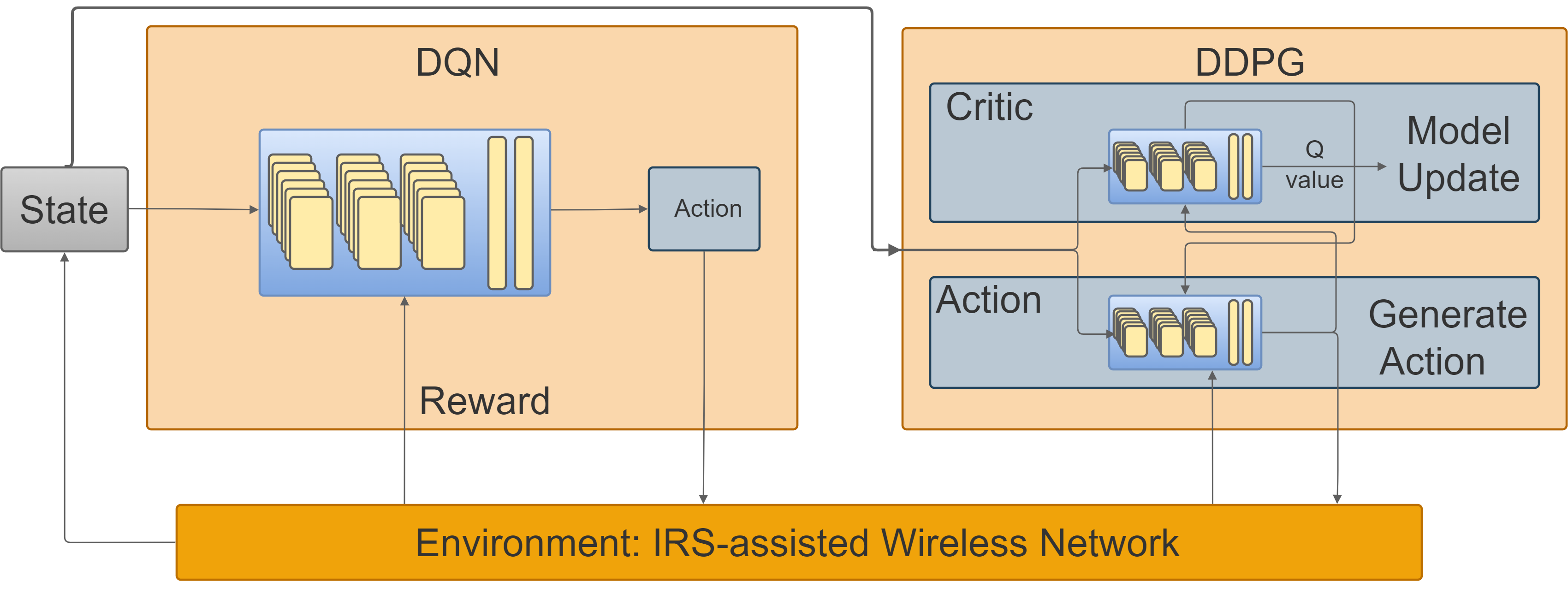} }
		\caption{In RL, the DQN and DDPG architectures accept same state (channel data and received SNR) and environment data (beamformers to be evaluated). The DQN involves training a single neural network based on the reward determined from the environment. On the other hand, the DDPG has multiple neural networks, where actor-critic architectures are used to compute actions and target values, respectively \citep{elbir2020survey}. 
		}
		\label{fig_DQN_DDPG}
	\end{figure}

	\section{Challenges and Future Outlook}
	\label{sec:pathForward}
	The techniques for RIS inverse design and processing are constantly evolving. Major challenges include reduction of training cost, gathering of labeled data, effective handling of system imperfections, and better data representations.
	
	\subsection{Design}
	New approaches are needed to increase the computational efficiency and reduce the amount of training required for DL-based RIS design. As mentioned below, reduction of design time and achieving full EM-compliance remains a major challenge.
	\subsubsection{Hybrid physics-based models}
	Hybrid models, where training set is supplemented by physics-based analytical models, reduce the amount of required training data and increase learning efficiency. Analytical RF circuit-based models are available to predict the performance of several canonical meta-atom designs. To speed up the training data generation, these analytical circuit-based models could be used to supplement the training data set and reduce iterations of time-consuming full-wave EM simulations. 
	It may also be feasible to create innovative DL design and optimization architectures that utilize physics-based analytical models within the ANN architecture. Another method to reduce the amount of required training data for multi-layer MTS designs is to use T-matrix data to analytically cascaded MTS designs from single-layer training data.
	
	\subsubsection{Other learning techniques}
	TL may also be used for expediting and improving the learning of a new task by using a previously trained neural network weights and bias as the initialization for the new ANN. Since all ANNs for meta-atom performance prediction and inverse design are implicitly learning Maxwell's equations, it is sensible that a network trained for one meta-atom design or frequency band is scaled and transferred to a related design. DQNs have also been studied to increase the efficiency of MTS holograms and automated multi-layer RIS design.
	
	\subsubsection{Improved data representation}
	More complex input data structures and representations are increasingly studied for DL-based RIS. While this article focused on discrete input parameters and image data structures are RIS design representations,  graphical and sequential data structures have recently been proposed as alternatives. The graphical model has been used to represent EM systems with near-field coupling (as in coupled resonators). In this arrangement, graph nodes contain resonator attributes, such as material, geometry, and location, and graph nodes represent the near-field coupling factors. These graphical data structures are processed using graphical neural networks (GNNs). While yet to be extensively explored in the domain of MTS, GNNs have been applied to model a broad range of physical systems. GNNs have the potential to handle additional complexities to jointly optimize RIS design and operation in wireless communication networks. 
	Additionally, sequential data structures are another data representation that is yet to be extensively explored in the context of MTS. 
	Similarly, sequential data structures are useful for representing time-sequence data in dynamic EM systems (as in RIS filters) and are learned using recurrent neural networks (RNNs). In other domains, such as natural language processing (NLP), sequential data is often learned using RNNs, which are ANNs that use forward or backward connections to enable a memory of internal states between successive passes to the network. As dynamic operation of RIS becomes increasingly important in the development of wireless networks and SRE, it is likely that RNNs will become increasingly useful to model dynamic RIS.
	
	subsection{Deep Reinforcement Learning (DRL)}
	Similar to evolutionary optimization techniques, RL is an area of ML concerned with how software agents ought to take actions in an environment in order to maximize the notion of cumulative reward through trial and error. Without the use of labeled training data, RL algorithms learn system dynamics through exploration to maximize a reward function. Here, DRL algorithms, such as DQNs, have produced ML advances in a broad range of applications including robotics, strategy games, NLP, and computer vision. To date, DQNs have been studied to increase the efficiency of MTS holograms and automated multi-layer MTS design. However, research using D-RLs for MTS design and optimization is very limited and further research is needed to develop these techniques for MTS applications. D-RL-based networks hold the promise for automated self-learning of RIS in SRE that are able to adapt and optimize themselves for dynamic RF environments and modulations.

	\subsection{Applications}
	Several challenges remain for DL architectures to reach their full potential in realizing significant performance gains and efficiency for RIS-assisted wireless systems. Given that it is an emerging technology, larger sets of real data are not yet available. Then, model training consumes much time and resources, including parallel processing and storage. Further, to achieve commercial viability of DL-based RIS-aided communications, dynamically adapting to changes in the environment is crucial. Finally, new RIS-specific implementation challenges have also been identified within emerging technologies such as terahertz communications, cell-free massive MIMO, drone operations, and open radio access network (RAN).
	
	\subsection{Channel Modeling}
	Channel Modeling is a challenging task in communication systems, especially with the large number of antennas due to the complexity of system architecture. In order to provide a reliable channel modeling performance, DL can be of help to construct a data-driven model based on the field measurements. In this case, SL schemes can be used to construct the channel model as a relationship between the input and output of a learning model~\citep{elbir_LIS}. Thus, DL-based methods are expected to become more frequently used in RIS-assisted wireless networks for channel modeling.
	
	\subsubsection{Data Collection}
	Massive data collection hampers successful performance of DL-based techniques for all wireless communications tasks: signal detection, channel estimation, and beamformer design. The  signal detection requires collection and storage of transmit and receive data symbols for different channel conditions. The prerequisites for channel estimation and beamforming are even more tedious because of additional labeling process. This is difficult to overcome in, especially online scenarios. Apart from SL, the label-free structure in RL is particularly helpful but at the cost of training times. It is possible to relax the data collection requirements by realizing the propagation environment in a numerical electromagnetic (EM) simulation tool~\citep{elbir2020_FL_CE} and then using a more realistically simulated data. This is helpful in constructing the training dataset offline but chances of failure remain in a real world scenario. Very recently, public datasets for channel estimation problem in RIS-aided communications were made available in the 2021 IEEE SIgnal Processing Cup competition.

	\subsubsection{Model Training}
	The models are usually trained offline prior to their online deployment at a PS connected to the BS. In addition, the model training complexity increases with the number of RIS elements and number of RISs deployed between the users and the BS. 	This introduces huge transmission overhead for model training. 	The FL has potential to reduce this cost and enable a communication-efficient model training (see, e.g., Fig.~\ref{fig_NMSE_ALL}). Here, combining the label-free structure of RL and the communications efficiency of FL, i.e. federated reinforcement learning, could be the next step.

	\subsubsection{Environment Adaptation and Robustness}
	The behavior of the channel affects all DL-based tasks including channel estimation, beamforming, user scheduling, power allocation, and antenna selection/switching. Addressing the trade-off between the bias and the variance of the model output is essential for robust performance. This is usually achieved using a validation data so that the learning model does not either over-fit or under-fit the training data. Nonetheless, this does not generalize the learning performance to different environments. Moreover, the current DL architectures for wireless systems remain environment-specific because the input data space of their learning model is limited. As a result, the performance degrades significantly when the learning model is fed with the input from unlearned/uncovered data space. In order to cover larger data spaces and provide a robust performance against the changes in the environment, wider and deeper learning models are required. But the current DL architectures for wireless communications comprise less than a million neurons and and are composed of only a few layers (Table 1)~\citep{elbir2020_FL_CE}. The giant learning models for image recognition or natural language processing consists of millions and billions of neurons, e.g., VGG (138 million), AlexNet (60 million), and GPT-3 (170 billion). Clearly, going wider and deeper in designing the learning models is of great interest for future DL-based RIS-aided systems.

	\section{Summary}
	\label{sec:summ}
	We surveyed DL-based techniques for designing RIS hardware to be deployed for future wireless communications. When the design space and scale of the RIS arrays increases, learning-based architectures outperform evolutionary optimization techniques for both surrogate performance modeling and inverse design. The DL inverse design is flexible in admitting a variety of RIS unit structures. The DGMs are the most useful because of their ability to generate new designs not previously seen in the published literature. While active research and techniques in this area are still evolving, DL is a promising solution for the inverse design of RIS.
	
	We also investigated DL architectures for RIS-assisted wireless systems for key applications of signal detection, channel estimation, and beamforming. We extensively discussed various learning schemes and model architectures, such as SL, UL, FL and RL for RIS applications. The SL exhibits better performance than UL and RL because of label usage. The UL and RL are label-free schemes that provide less complexity during training data generation. However, UL still involves an optimization stage for each data instance. Among all, the RL is the most promising technique because of its standalone operation and the consequent ability to adapt to environmental changes at the cost of longer training times.
	
	The FL reduces the transmission overhead significantly and can be integrated with the other learning methods. The combination of FL- and RL-based learning policies not only exhibits a communication-efficient model training but also provides environmental adaptation. Major research challenges include data collection, model training, and environment adaptation. These should be addressed simultaneously to provide a reliable DL architecture for the next-generation RIS-assisted wireless systems. Specifically, the combination of FL and RL should be fed with the collection of huge datasets and massive neural networks so that a robust DL architecture is achieved.

    \section*{Acknowledgement}
    The authors warmly acknowledge valuable contributions of Dr. John A. Hodge (Amazon) for the inverse design portion of this chapter, when he was a graduate student at Virginia Tech.

\bibliography{main}%

\latexprintindex

\end{document}